\documentclass[
 prb,
 reprint,
 twocolumn,
 superscriptaddress,
 noeprint
]{revtex4-2}
\usepackage[]{standalone}
\usepackage{import}
\usepackage[utf8]{inputenc}
\usepackage[T1]{fontenc}
\usepackage{cmap}
\usepackage{graphicx}
\usepackage[hyperfootnotes=false,breaklinks=true]{hyperref}
\usepackage{setspace}
\hypersetup{
 bookmarksopen=true,
 bookmarksopenlevel=1,
 colorlinks=true,
 linkcolor=blue,
 anchorcolor=blue,
 citecolor=blue,
 filecolor=blue,
 urlcolor=blue,
 pdfpagemode=UseOutlines,
 pdfstartview={XYZ null null 1},
 linktocpage=true,
}
\usepackage{amsmath}
\usepackage{amssymb}
\usepackage{amstext}
\usepackage{amsfonts}
\usepackage{amsthm}
\usepackage{mathtools}
\DeclarePairedDelimiter\ceil{\lceil}{\rceil}
\DeclarePairedDelimiter\floor{\lfloor}{\rfloor}
\usepackage{bm}
\usepackage{braket}
\usepackage{physics}
\usepackage{caption}
\usepackage{subcaption}
\usepackage[
 capitalise,
]{cleveref}
\crefname{table}{Tab.}{Tabs.}%
\crefname{section}{Sec.}{Sects.}%
\Crefname{section}{Section}{Sections}%
\Crefname{table}{Table}{Tables}%
\usepackage{enumitem}

\usepackage{tikz}
\usetikzlibrary{external}
\usetikzlibrary{
  shapes.geometric,
  arrows,
  arrows.meta,
  positioning,
  backgrounds,
  fit,
  intersections,
}
\usepackage{qcircuit}
\xyoption{all}
\xyoption{pdf}

\newcommand{\ra}{\rangle}
\newcommand{\la}{\langle}
\newcommand{\ua}{\uparrow}
\newcommand{\da}{\downarrow}

\newcommand{\nint}{\operatorname{NINT}}
\newcommand{\pdag}{{\phantom{\dagger}}}
\newcommand{\ibmqManila}{\textit{ibmq\_\kern-0.2em manila}}

\makeatletter
\def\blfootnote{\xdef\@thefnmark{}\@footnotetext}
\makeatother

\newcommand{\mycomment}[1]{}

\usepackage[draft, commentmarkup=uwave]{changes}
\definechangesauthor[name={Thomas Steckmann}, color=red]{TS}
\definechangesauthor[name={Trevor Keen}, color=orange]{TK}
\definechangesauthor[name={Alexander F. Kemper}, color=blue]{Lex}
\definechangesauthor[name={Efekan K\"{o}kc\"{u}}, color=purple]{EK}
\definechangesauthor[name={Eugene F. Dumitrescu}, color=green]{EFD}
\definechangesauthor[name={Yan Wang}, color=cyan]{YW}

\graphicspath{{figs/}{texfigs/}}

\begin{document}

\title{Mapping the metal-insulator phase diagram by algebraically fast-forwarding dynamics on a cloud quantum computer}

\author{Thomas Steckmann}
\email{tmsteckm@umd.edu}
\altaffiliation{present address: Joint Center for Quantum Information and Computer Science, NIST/University of Maryland.}
\affiliation{{Department of Physics, North Carolina State University, %
Raleigh, North Carolina 27695, USA}}

\author{Trevor Keen}
\affiliation{{Department of Physics and Astronomy, University of Tennessee, %
Knoxville, Tennessee 37996}}

\author{Efekan K\"{o}kc\"{u}}
\affiliation{{Department of Physics, North Carolina State University, %
Raleigh, North Carolina 27695, USA}}

\author{Alexander~F.~Kemper}
\affiliation{{Department of Physics, North Carolina State University, %
Raleigh, North Carolina 27695, USA}}

\author{Eugene F.~Dumitrescu}
\affiliation{{Quantum Computational Science Group, Oak Ridge National Laboratory, %
Oak Ridge, Tennessee 37831, USA}}

\author{Yan Wang}
\email{wangy2@ornl.gov}
\affiliation{{Quantum Computational Science Group, %
Oak Ridge National Laboratory, %
Oak Ridge, Tennessee 37831, USA}}

\begin{abstract}
Dynamical mean-field theory (DMFT) maps the local Green's function of the Hubbard model to that of the Anderson impurity model and thus gives an approximate solution of the Hubbard model from the solution of simpler quantum impurity model. Accurate solutions to the Anderson impurity model nonetheless become intractable for large systems. Quantum and hybrid quantum-classical algorithms have been proposed to efficiently solve impurity models by preparing and evolving the ground state under the impurity Hamiltonian on a quantum computer that is assumed to have the scalability and accuracy far beyond the current state-of-the-art quantum hardware. As a proof of principle demonstration targeting the Anderson impurity model we, for the first time, close the DMFT loop with current \emph{noisy} hardware. With a highly optimized fast-forwarding quantum circuit and a noise resilient spectral analysis we observe a Mott phase transition. Based on a Cartan decomposition, our algorithm gives a fixed depth, fast-forwarding, quantum circuit that can evolve the initial state over \emph{arbitrarily long} times without time-discretization errors typical of other product decomposition formulas such as Trotter decomposition. By exploiting the structure of the fast-forwarding circuits we reduce the gate count (to 77 CNOTs after optimization), simulate the dynamics, and extract frequencies from the Anderson impurity model on noisy quantum hardware. We then demonstrate the Mott transition by mapping the full metal-insulator phase-diagram. Near the Mott phase transition, our method maintains accuracy where Trotter error would otherwise dominate due to the long-time evolution required to resolve quasiparticle resonance frequency extremely close to zero. This work presents the first computation of the Mott phase transition using noisy digital quantum hardware, made viable by a highly optimized computation in terms of gate depth, simulation error, and runtime on quantum hardware. 
To inform future computations we analyze the accuracy of our method vs. a noisy Trotter evolution in the time-domain.  Both algebraic circuit decompositions and error mitigation techniques adopted could be applied in an attempt to solve other correlated electronic phenomena beyond DMFT on noisy quantum computers.
\end{abstract}

\date{\today}

\blfootnote{This manuscript has been authored by UT-Battelle, LLC, under Contract No.~DE-AC0500OR22725 with the U.S.~Department of Energy. The United States Government retains and the publisher, by accepting the article for publication, acknowledges that the United States Government retains a non-exclusive, paid-up, irrevocable, world-wide license to publish or reproduce the published form of this manuscript, or allow others to do so, for the United States Government purposes. The Department of Energy will provide public access to these results of federally sponsored research in accordance with the DOE Public Access Plan.}

\maketitle

\section{Introduction}
Using quantum computers to accurately model the behavior of strongly correlated many-body quantum systems is one of the most promising near-term applications of noisy intermediate scale quantum (NISQ) computers. For example, quantum simulations of fermions only require ${\sim} 100$ data qubits to potentially surpass classical simulation methods. In contrast, Shor's algorithm~\cite{Shor1997} for factoring large numbers will require tens of thousands of logical qubits comprising tens of millions of physical qubits to become practically useful~\cite{Beverland2022}. 
A plethora of proposals for simulating correlated fermionic systems using quantum computers exist~\cite{Bauer2016, Kivlichan2018, Jiang2018, Wecker2015, Cade2020, bassman2021simulating, Jamet2021}, but relatively few have been implemented or tested on current \emph{noisy} devices~\cite{Linke2018,Rungger2019, keen_quantum-classical_2020, Backes2023}.

A wide variety of paradigmatic correlated condensed-matter systems can be mapped to a simpler, corresponding quantum impurity model by embedding methods such as dynamical mean-field theory (DMFT). Despite the simplification, classical simulations based on exact diagonalization (ED) are still limited to dozens of orbitals \cite{Liebsch2012} due to the exponential growth of Hilbert space. Other methods, such as quantum Monte Carlo (QMC)~\cite{GullQMC2011} and matrix product state (MPS)~\cite{WolfMPS2014} methods, also suffer from some sort of exponential complexity scaling making them intractable. Specifically, in the case of QMC, the fermion minus sign problem~\cite{Troyersign2005} has been shown to be NP-hard in general and limits effective simulations to only high temperatures. The exponential scaling of MPS is due to entanglement issues for certain geometries~\cite{WolfMPS2014}. Quantum computers alleviate the exponential scaling by instead storing many-body quantum states with a quantum memory resource that scales polynomially with the system size.

In this work, we solve the two-site DMFT of the archetypal Hubbard model by utilizing a Lie-algebraic method to fast-forward the dynamics of the corresponding Anderson impurity model (AIM). Our fast-forwarding method, based on a Cartan decomposition of the algebraic closure of the AIM Hamiltonian, compiles the time evolution operator of AIM Hamiltonian into a fixed depth circuit for \emph{any} chosen evolution time $t$. Therefore, the error from the quantum device is independent of $t$ and a one-time decomposition performed on a classical computer allows for arbitrarily low numerical error in the parameters of the decomposed factors. As described in Ref.~\onlinecite{kokcu_fixed_2021}, given a Hamiltonian $\hat{H}$, the Cartan decomposition requires finding a particular sequence of unitary rotations which, when contracted, span and parameterize the time evolution unitary $e^{-it\hat{H}}$ of a target system for all time $t$. The Cartan decomposition generalizes the polar and singular-value decompositions at the level of groups and provides a mapping from the required dynamics onto two sets of parameterized unitaries.

We first review and motivate the quantum impurity model and provide our hybrid quantum-classical algorithm solving it. Next, we apply group analysis to fast-forward the dynamical simulation on a quantum computer. The structure of Cartan decomposition allows for further optimizing the time evolution circuits to tailor to specific hardware architectures. Finally, we perform simulations at vastly different time scales and apply Fourier transform spectrum analysis to accurately extract both low- and high-frequency poles from accordingly sampled Green's functions for the two-site DMFT on NISQ hardware. This enables us to demonstrate for the first time a Mott-insulating phase transition in the Hubbard model via a digital quantum simulation on quantum hardware.

\section{Quantum Algorithms for Hamiltonian Simulation and DMFT}
In this section we briefly review a variety of existing quantum algorithms for the DMFT problem, although none of them is particularly successful in solving this problem. Our algorithm will be detailed in \cref{sec:gfAlg} including the fast-forwarding algorithm in \cref{sec:cartan}, and future directions are discussed in \cref{sec:future_directions}.

In DMFT one must self-consistently solve the electron Green's function (also known as response function or propagator) for the corresponding Anderson impurity model. On a quantum computer, this quantity can be measured via \emph{Hamiltonian simulation}, which broadly refers to approximately compiling the time evolution operator $U(t) = {\exp}(-i\hat{H}t)$ into a sequence of physically realizable unitary operators, i.e., digital quantum gates. A wide variety of advanced Hamiltonian simulation algorithms exist, each having a computational runtime determined primarily by the scaling of the approximation in terms of the simulation time $t$, system size $N$, and desired approximation error $\varepsilon$. However, in practice these algorithms assume quantum hardware with arbitrarily small physical error rates. This is at odds with simulations run on near-term hardware where the results severely depend on the physical error rates and therefore require significant overhead in the form of, e.g. calibrations, error-processing, filtering, and additional noise-reducing characterizations. As discussed next, there is to date no general algorithm which optimizes across all scales, e.g., having optimal runtime scaling while remaining suitable for current noisy experimental quantum computing platforms.


One class of Hamiltonian simulation algorithms are termed ``asymptotic,'' meaning that they
have already almost saturated the expected optimal scalings, e.g., linear scaling in the system's interaction strengths in terms of $\norm*{\hat{H}}$, linear scaling in the simulation time $t$, and $\varepsilon^{-1}$ or $\log \varepsilon^{-1}$ scaling in the desired approximation error $\varepsilon$~\cite{Low2017}. Despite realizing Feynman's vision of natural asymptotic scaling, these methods' utility is severely restricted in the near term due to their reliance on the assumption of arbitrarily high accuracy quantum operations which cannot be achieved in the general case outside of the era of large-scale, fault-tolerant quantum computers (FTQC). In contrast, promising near-term algorithms such as low-order Trotter-Suzuki product formulas and variational methods avoid some of the high overhead costs associated with asymptotic algorithms, such as those arising from requiring large registers of ancilla control qubits or higher-order product formula expansions, yet these near-term algorithms still face a significant challenge with \textit{long-time} simulation due to the accumulated error introduced by relatively large number of physical circuit operations on current noisy quantum hardware.

Prior work studying the dynamics of interacting electrons on current quantum hardware via DMFT observed that even over very small time scales Trotter-based approximate time evolution lead to nonphysical results: compared to theoretical values, simulations on the quantum computer give inaccurate frequencies in the time evolution for the two-site DMFT, which are symptoms of decoherence or approximation errors~\cite{keen_quantum-classical_2020}. For NISQ systems, the Trotter approximation leads to a dilemma: theoretically it becomes exact in the limit of an infinite-depth circuit, so more accurate simulations require increased gate counts, but increasing gate count reduces simulation fidelity due to accumulated noise. The hardware requirements needed to achieve reliable updates in the DMFT loop using the Trotter decomposition of the time evolution operator has been analyzed in Ref.~\onlinecite{Jaderberg2020} in terms of the CNOT two-qubit gate fidelity: to achieve perfect agreement with the exact solution, first-order Trotter based simulation requires a fidelity of $\mathcal{F}_{\text{CNOT}} > 99.999\%$, or $\mathcal{F}_{\text{CNOT}} > 99.9\%$ after applying a variational recompilation algorithm (termed incremental structural learning by those authors). 

In an alternate approach~\cite{Rungger2019} without direct Hamiltonian simulation, the authors use a variational quantum eigensolver (VQE) to implement an exact diagonalization solver for the two-site DMFT problem. This method works well for two-site DMFT after a regularization technique is used to remove the unphysical pole that arises from small errors, but the usefulness of the method depends on the scalability of VQE that is hindered by the classical optimization part of this hybrid algorithm~\cite{Cerezo2021, Tilly2022, Bittel2021} and the need to resolve an exponentially growing number of eigenenergies with increasing system size.
Another potential approach focuses on error mitigation techniques which trade runtime for accuracy by taking, in general, exponentially many additional measurements of the system to reconstruct the noise-free operations~\cite{Czarnik2021}. These methods' complexity is hidden in the exponential growth of the mitigation technique's sample complexity as one extracts an infinitesimally small signal from a noisy quantum computer~\cite{Quek2022}. 
Thus, despite years of work on this problem a \emph{reliable} NISQ-friendly algorithm that enables the closure of the DMFT loop has not been achieved. 

These issues illustrate a need for new approaches that can extract meaningful information from fragile NISQ simulations. In \cref{sec:gfAlg}, we describe our new hybrid DMFT algorithm that uses a fast-forwarding algorithm (\cref{sec:cartan}) to compress a Hamiltonian simulation circuit's depth further than the asymptotic linear scaling. For large interacting systems with arbitrary Hamiltonians this would be prohibitively expensive due to the no-fast-forwarding theorem. (For free-fermion noninteracting Hamiltonians of any system size, our fast-forwarding algorithm always matches the optimal asymptotic scaling from Ref.~\onlinecite{Gu2021}.)
However, for small interacting systems such as the two-site AIM considered in this work, the overhead of our fast-forwarding algorithm---although would be exponential in increasing bath sites---remains manageable and allows for simulating arbitrarily long time scales with constant gate depth. To offset this overhead scaling and close the DMFT loop, we take advantage of the structure of Cartan decomposition to further shorten the circuit (\cref{sec:hardwareCirc}), use randomized Cartan solutions to mitigate coherent noise in additional to other error mitigation techniques (\cref{sec:errorMit}), and apply Fourier transform to accurately extract the Green's function frequencies from noisy data (\cref{sec:freqExtra}).

\section{Model Hamiltonians} \label{sec:model}

\subsection{Hubbard model} \label{sec:modelHubbard}
The Fermi-Hubbard (abbreviated as Hubbard below) model has no known exact solution except in one and infinite dimensions, even though it is one of the simplest models for interacting electrons. Despite its deceptive simplicity, the Hubbard model can account for many interesting strongly correlated quantum phenomena in condensed matter physics, including the Mott metal-insulator transition~\cite{Mott1968, Imada1998, Georges1992}, antiferromagnetism~\cite{White1989}, emergent spin and stripe orders~\cite{Zheng1155, Huang2018}, strange metallic behavior~\cite{Huang987}, pseudogaps~\cite{Gull2013, Chen2015}, and high-temperature superconductivity~\cite{Maier2005,Gull2013}.

The single-band Hubbard model Hamiltonian~\cite{hubbard_electron_1963} is given by
\begin{align}
     \hat{H}_\text{Hub}
  =& -\tilde{t} \sum_{\la i,j\ra, \sigma}
     (\hat{c}_{i\sigma}^\dagger \hat{c}_{j\sigma}^\pdag +
       \hat{c}_{j\sigma}^\dagger \hat{c}_{i\sigma}^\pdag) +
     U \sum_i \hat{n}_{i\ua} \hat{n}_{i\da} \notag\\& -
     \mu \sum_{i,\sigma} \hat{n}_{i\sigma},
  \label{eq:HubbardHam}
\end{align}
where $\la i,j \ra$ denotes nearest-neighbor sites $i$ and $j$, $\hat{c} _{i\sigma}^\dagger $ ($\hat{c} _{i\sigma}^{\phantom{\dagger}}$) is the electron creation (annihilation) operator for electron with spin $\sigma\in \{\ua, \da\}$ at lattice site $i$, $\hat{n} _{i\sigma} = \hat{c} _{i\sigma}^\dagger \hat{c} _{i\sigma}^{\phantom{\dagger}}$ is the electron density operator, $\tilde{t}$ is the hopping integral (tunneling), $U > 0$ is the local (on-site) Coulomb interaction, and $\mu$ is the chemical potential. In the context of quantum computing, the Hubbard model has been recently investigated with the applications of VQE algorithms~\cite{Wecker2015vqe, Reiner2019} and as a benchmark for quantum simulations~\cite{Jiang2018, Kivlichan2020, Wecker2015}.

\subsection{Anderson impurity model and dynamical mean-field theory} \label{sec:modelAIMdmft}
Simulations of the Hubbard model are limited to dozens of fermionic orbitals~\cite{Ehlers2017, Venderley2019, Lunts2021}, far from the large number of particles in the macroscopic (thermodynamic) limit. DMFT~\cite{kotliar_strongly_2004} is a significant development in studying the Hubbard model in the thermodynamic limit. In the infinite spatial dimension ($\infty$-$d$) limit, such as $\infty$-$d$ hypercubic lattice and Bethe lattice with infinite coordination number, DMFT \emph{exactly} maps the solution of the Hubbard model to that of the AIM, in the sense that the temporal correlations are accurately captured. Specifically, in DMFT the interacting electrons in the Hubbard model are reduced to electrons interacting on a single impurity site coupled to a noninteracting electronic bath of \emph{continuous} levels that tunnel into the impurity site. In practice the levels are often approximated by $N_b$ discrete bath sites with on-site energy $\epsilon_i$ and index $i\in \{1,\dots,N_b\} \equiv [N_b] $. When $N_b = \infty$ (i.e., the infinite dimension limit), the DMFT solution to the Hubbard model becomes exact. The AIM Hamiltonian is given by
\begin{align}
    \hat{H}_\text{AIM}
 =& \sum_{i = 1,\sigma}^{i = N_b} V_i(
      \hat{c}^\dagger_{0,\sigma} \hat{c}^\pdag_{i,\sigma} +
      \hat{c}^\dagger_{i,\sigma} \hat{c}^\pdag_{0,\sigma}) + 
     U \hat{n}_{0,\uparrow} \hat{n}_{0,\downarrow} \notag \\ &+
     \sum_{i = 0, \sigma}^{i = N_b} (\epsilon_i - \mu) \hat{n}_{i,\sigma}, 
 \label{eq:AIMham}
\end{align}
where the hybridization parameter $V_i$ is the hopping between the impurity site (site-index $i=0$) and bath sites (site-index $ i \in  [N_b]$). The Coulomb interaction $U$-term only involves the impurity site. Since we will consider nonmagnetic states, $V_i$ and $\epsilon_i$ do not depend on the electron spin $\sigma$.

The minimal realization of DMFT for the Hubbard model dynamics is the two-site DMFT~\cite{potthoff_two-site_2001}, involving the impurity site and only one bath site ($N_b = 1$). In this work we will consider solving this case on NISQ hardware. After the Jordan-Wigner fermion-spin transform (see \cref{sec:JW}), \cref{eq:AIMham} in fermion operators becomes the impurity Hamiltonian in Pauli string operators requiring $2(N_b+1)$ qubits and given by
\begin{align}
    \hat{H}_\text{AIM} 
 &= \frac{V}{2}(X_0X_1 + Y_0Y_1 + X_2X_3 + Y_2Y_3) + \frac{U}{4}Z_0Z_2. \label{eq:AIM2Site}
\end{align}
for the $N_b = 1$ case. Here, $X_l$, $Y_l$, and $Z_l$ are Pauli operators acting on qubit $l$. Specifically, the spin $\ua$ and $\da$ modes on the impurity site $i=0$ (the bath site $i=1$) are mapped to qubit $0$ and $2$, respectively (qubit $1$ and $3$, respectively). In addition, we only consider the half-filled paramagnetic ground state, so the impurity Hamiltonian \cref{eq:AIM2Site} has been simplified by setting $\mu = \frac{U}{2}$, $\epsilon_0 = 0$, and $\epsilon_1 = \frac{U}{2}$ in \cref{eq:AIMham}.

\section{Algorithm for computing Green's function}\label{sec:gfAlg}
In DMFT, the dynamic response of the interacting electron system is described by the \emph{retarded impurity Green's function} denoted as $G^{R,\ua}_{\text{imp}}(t, t')$ (for spin-up orbital on the impurity site) in the time domain and is given by
\begin{align}
    G^{R,\ua}_\text{imp} (t, t') 
 &= -i\theta(t-t')
     \ev**{\qty{\hat{c}_{0}^\pdag (t), 
       \hat{c}_{0}^\dagger (t')}}
     {\psi_0},
 \label{eq:Gimpta}
\end{align}
where $\theta(t)$ is the step-function, $\hat{c}_{0}(t) = U^\dagger(t) \hat{c}_{0} U(t)$, $\hat{c}_{0}^\dagger(t) = U^\dagger(t) \hat{c}_{0}^\dagger U(t)$, time evolution operator $ U(t) = e^{-it\hat{H}_\text{AIM}}$, and $\ket{\psi_0}$ is the many-body ground state of the impurity Hamiltonian $\hat{H}_\text{AIM}$ (the subscript ``$0$'' in $\ket{\psi_0}$ indicates ground state, not site index $0$). Due to time-translation invariance of a time-independent Hamiltonian $\hat{H}_\text{AIM}$, we simplify the computation by setting $t' = 0$. Since we only consider the paramagnetic ground state, the impurity Green's function is diagonal in spin-space $G^{R,\sigma\sigma'}_\text{imp}(t) = G^{R,\sigma}_\text{imp}(t)\delta_{\sigma\sigma'}$ and $G^{R,\ua}_\text{imp}(t) = G^{R,\da}_\text{imp}(t)$. We therefore drop the spin index in the remainder of this work and denote the impurity Green's function by $G^\text{R}_{\text{imp}} (t)$.

In \cref{sec:GFeval} we elaborate on the full expansion and subsequent simplification of Green's function after the Jordan-Wigner transform is applied, which results in the relatively simple expression
\begin{align}
    iG^\text{R}_{\text{imp}} (t>0)
 &= \Re \ev**{U^\dagger(t) X_0 U(t) X_0}{\psi_0}
  \label{eq:reducedGreens}
\end{align}
that is easy and inexpensive to measure on quantum hardware. This term can be measured with a single Hadamard-test type quantum circuit as shown in \cref{fig:Circuit Elements}(b) using only a single time evolution unitary as discussed \cref{sec:hardwareImpl}.

\subsection{Iteration loop for DMFT}\label{sec:dmftLoopAlg}
The DMFT mapping is a self-consistent mapping, requiring multiple iterations where the AIM Hamiltonian parameters $V_i$ and $\epsilon_i$ are updated, from an initial guess, until the system reaches self-consistency. At each new iteration, the parameters $V_i$ and $\epsilon_i$ computed in the previous iteration are put back into the impurity model, whose Green's function is then solved, and the solution is used to recompute these parameters. The iteration loop continues until the recomputed parameter values are sufficiently close to the previous ones. For the two-site model, particle-hole symmetry and the structure of the two-site solution provide a mechanism for reducing the cost of the computation and improving the accuracy of convergence.

\begin{figure*}
  \centering
  \setlength{\fboxsep}{0pt}
  \includegraphics{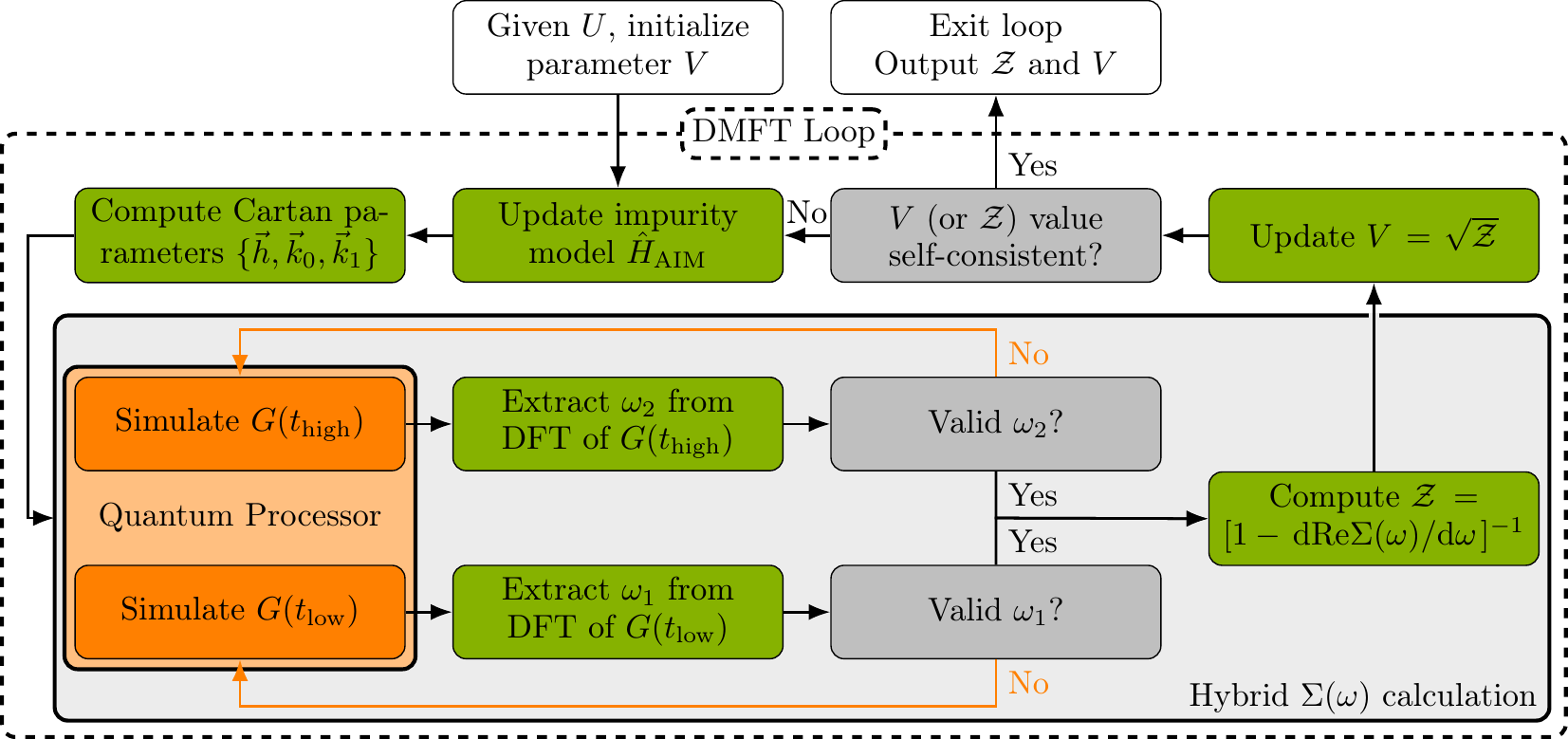}
\caption{Flow diagram of the DMFT loop specialized for the two-site calculation. Our calculations are initialized with $V = 0.5$. Each DMFT loop iteration also updates the time evolution Cartan parameters corresponding to the updated $V$, although the Hamiltonian algebra remains the same (so the group analysis needs to be done only once). The hybrid computation of $\Sigma(\omega)$ evaluates the two frequencies $\omega_1$ and $\omega_2$ separately, in a procedure that is elaborated on in \cref{sec:freqExtra}.}
  \label{fig:DMFT Loop}
\end{figure*}
The steps of DMFT loop used in our calculations are represented in \cref{fig:DMFT Loop} and summarized as follows.
\begin{enumerate}
  \item Choose initial values for parameters $V_i$ and $\epsilon_i$. Due to half-filling of the two-site model, the values for $\epsilon_0$ and $\epsilon_1$ are fixed and do not need to be updated.
  
  \item On the quantum computer, simulate $\hat{H}_\text{AIM}$, and then sample and measure $G^R_\text{imp}(t)$ for a selection of time $t$ values. In the two-site model, $G^R_\text{imp}(t)$ has the following analytical form
  \begin{align}
      iG^R_\text{imp}(t>0) 
   &= 2[\alpha_1 \cos{(\omega_1 t)} + \alpha_2 \cos{(\omega_2 t)}],
   \label{eq:exactGt}
  \end{align}
  where $\omega_1$ is the quasiparticle resonance frequency and $\omega_2$ corresponds to the Hubbard band~\cite{potthoff_two-site_2001}.
  
  \item Compute the discrete Fourier transform (DFT) of sampled time-domain Green's function, giving $G^R_\text{imp}(\omega)$.
    
  \item Compute the self-energy $\Sigma_\text{imp}(\omega)$.

  \item From the self-energy, compute the quasiparticle weight $\mathcal{Z}$ and update $V_\text{new} = \sqrt{\mathcal{Z}}$. In this work, we use the value of the quasiparticle weight computed using the derivative of the self-energy at zero frequency,
  \begin{align}
      \mathcal{Z}^{-1} 
   &= 1 -  \left. \dv{{\Re}[\Sigma_{\text{imp}}(\omega)]}{\omega} \right|_{\omega = 0}.
   \label{eq:quasiDef} 
  \end{align}
\end{enumerate}

In \cref{sec:Self-energy-derivative} we derive an analytical form of the derivative in \cref{eq:quasiDef} to avoid the numerical instability of evaluating $\Sigma_\text{imp}(\omega)$ and its derivative near $\omega = 0$, and also to remove dependence of amplitudes $\alpha_1$ and $\alpha_2$ whose values are too sensitive to the hardware noise. The final equation used to compute the quasiparticle weight is as follows:
\begin{align}
    \mathcal{Z} 
 &= \frac{\omega_1^2 \omega_2^2}{V^2(\omega_1^2 + \omega_2^2 - V^2)},
    \label{eq:ZderivativeOmegaOnly}
\end{align}
where $\omega_1$ and $\omega_2$ are extracted from the Fourier transform of the sampled time-domain Green's function using the method given in \cref{sec:freqExtra}, a method we find very robust in obtaining accurate frequencies from noisy data.

\subsection{Cartan decomposition}\label{sec:cartan}

\begin{figure*}
  \centering
  \includegraphics[width=0.9\linewidth]{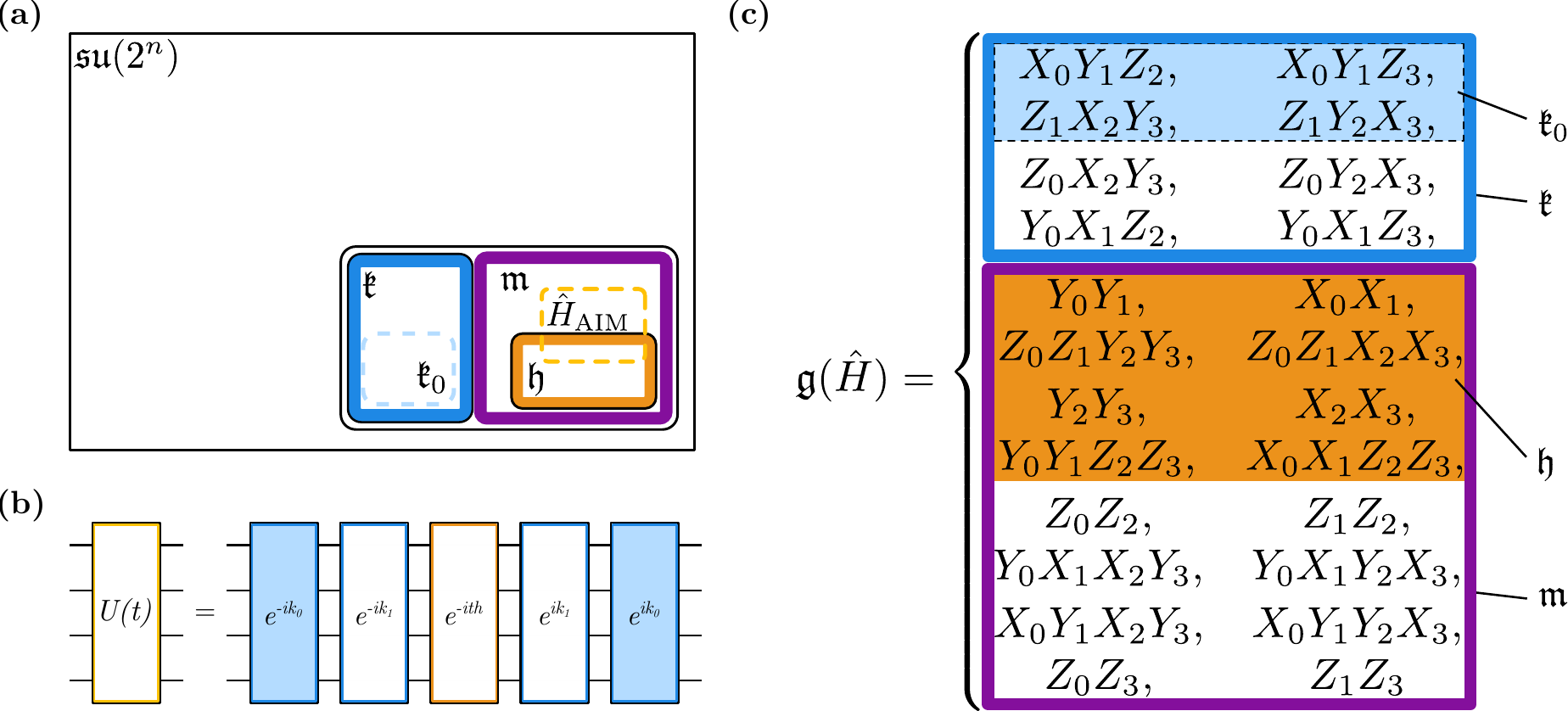}
\caption{(a) A generalized diagram of the Cartan decomposition of the Hamiltonian algebra with dimension = 24 within the special unitary algebra with dimension = 255. Here, $\mathfrak{k_0}$ is the set of basis elements which commute with $X_0$, which is not a typical requirement of Cartan decomposition but results in a significant gate cost reduction in our application. (b) A block circuit diagram of the decomposed time evolution operator. (c) Cartan decomposition  applied to the AIM Hamiltonian \cref{eq:AIMJWham}, where the blue, shaded light blue, magenta, and shaded orange color regions correspond to the sets $\mathfrak{k}, \mathfrak{k_0}, \mathfrak{m},$ and $\mathfrak{h}$.}
  \label{fig:AIM Decomposition}
\end{figure*}

The fast-forwarding algorithm used in our time evolution operator is based on the application of an algorithm for performing Cartan decomposition. Here, we briefly summarize the algorithm applied to $\hat{H}_\text{AIM}$ given in \cref{eq:AIM2Site} and also illustrate the steps in \cref{fig:AIM Decomposition}. We denote a (real) Lie algebra and its elements by lowercase Fraktur and Roman characters, respectively, such as $\mathfrak{g}$ and $ig\in \mathfrak{g}$, where $g$ is a $n$-qubit Pauli string or a linear combination of them (with real coefficients). Here, we use the physicists' convention with a prefactor $i$ in the Lie algebra elements. SU($2^n$) group elements are denoted by uppercase Roman characters, e.g., $G = \exp(ig)$.

The goal is to find a factorization of the time evolution unitary operator by use of the KHK theorem, which states that the unitary may be written as
\begin{equation}
    e^{-it\hat{H}} = e^{ik} e^{-ith} e^{-ik}
\end{equation}
where $k$ and $h$ are elements of a Cartan decomposition (see below). Note that the time argument $t$ only appears in one factor. The general steps to obtain the Cartan form of the time evolution operator are detailed in Refs.~\onlinecite{earp_constructive_2005,kokcu_fixed_2021}; we briefly summarize them here for completeness.
\newcounter{saveenum}
\begin{enumerate}
 \item Generate the \emph{Hamiltonian algebra} $\mathfrak{g}(\hat{H})$. This is a Lie algebra over the field $\mathbb{R}$ that is generated by the closure of commutators (Lie brackets) of $ib_l$, where $b_l$'s are individual $n$-qubit Pauli string terms of the Hamiltonian $\hat{H} = \sum_{l} \beta_l b_l$, ($\beta_l \in \mathbb{R}$). $\mathfrak{g}(\hat{H})$ a subalgebra of $\mathfrak{su}(2^n)$.

 \item Find a Cartan decomposition $\mathfrak{g} = \mathfrak{k} \oplus \mathfrak{m}$ of the Hamiltonian algebra $\mathfrak{g}(\hat H)$ such that $i\hat H$ lies in $\mathfrak{m}$. Here, $\mathfrak{k}$ is a subalgebra of $\mathfrak{g}(\hat H)$.

 \item From $\mathfrak{m}$ find a largest commuting subalgebra (i.e. a maximal Abelian subalgebra) $\mathfrak{h}$, which is called a \emph{Cartan subalgebra} of the pair $(\mathfrak{g}, \mathfrak{k})$.
 
 \item Find a local extremum over the algebra $\mathfrak{k}$ of $f(k) = \la e^{ik}(v)e^{-ik}, \hat{H} \ra$. Here, $\la a,b \ra$ is the Killing form proportional to $\Tr(ab)$ for $a,b\in \mathfrak{su}(2^n)$. $ik$ is an element of $\mathfrak{k}$ written as a sum of Pauli strings $k = \sum_j \kappa_j k_j$ where $ik_j$ form a basis for $\mathfrak{k}$. The optimization is performed over the coefficients $\kappa_j$. $v$ is a fixed element in $\mathfrak{h}$: $v = \sum_j \gamma^j h_j$ where $ih_j$ are Pauli strings that form a basis for $\mathfrak{h}$, and $\gamma$ is a transcendental number such as $\pi$. Here, $\gamma^j$ is the $j$-th power of $\gamma$.

 \item Compute the vector $e^{-ik}(i\hat{H})e^{ik} = ih$.
 \setcounter{saveenum}{\value{enumi}}
\end{enumerate}

The results of the algorithm are the elements $ih \in \mathfrak{h}$ and $ik \in \mathfrak{k}$ which satisfy $e^{-it\hat{H}} = e^{ik}e^{-ith}e^{-ik}$. Often, additional decomposition is required to implement $e^{ik}$ using a universal gate set including only single-qubit and CNOT gates, but in the case of two-site DMFT $\mathfrak{k}$ is Abelian and the decomposition is straightforward~\footnote{The algorithm from Ref.~\onlinecite{kokcu_fixed_2021} and the accompanied code~\cite{steckmann_cartan_2021} we actually use in our calculation removes additional decomposition in the non-Abelian cases by directly using the product form $\prod_j e^{i \tilde{\kappa}_j k_j}$ instead of $e^{ik} = e^{i\sum_j \kappa_j k_j}$ in the KHK Cartan decomposition, which is always possible since the Pauli string $k_j$'s constitute a full basis for $\mathfrak{k}$.}.
Because $h$ is always composed of commuting elements, the full exponential is relatively simple to implement exactly on a quantum computer. 

We note that the dimensionality of the Hamiltonian algebra generated by the $\hat{H}_\text{AIM}$ scales exponentially with the number of bath sites. However, for the two-site model, the size of the algebra remains manageable. It is an open question of continuing interest if the dimensionality of the Hamiltonian algebra can be constrained to polynomial in the number of bath sites by adopting some effective approximate algorithm. On the other hand,  for the ground state energy problem and ground state preparation problem, there exists an approximate algorithm with polynomial (in bath sites) runtime \cite{Bravyi2017}.

Analysis of the terms in $\mathfrak{k}$ resulting from the Cartan decomposition in \cref{fig:AIM Decomposition}(c) reveals that we can divide $\mathfrak{k}$ into a set of basis elements which commute with $X_0$, which we call $\mathfrak{k}_0$, and the elements which do not, which we call $\mathfrak{k}_1$. 
\begin{enumerate}
  \setcounter{enumi}{\value{saveenum}}
  \item Decompose $\mathfrak{k}$ into $\mathfrak{k}_0$ and $\mathfrak{k}_1$ such that $\mathfrak{k} = \mathfrak{k}_0 \oplus \mathfrak{k}_1$ and $[\mathfrak{k}_0,X_0] = 0$.
\end{enumerate}
This step later leads to a reduction in the circuit construction (see \cref{sec:hardwareCirc}), but we highlight the useful partition here due to the flexible product form of the Cartan decomposed time evolution operator.


\begin{figure*}[ht]
  \centering
  \includegraphics[scale=1]{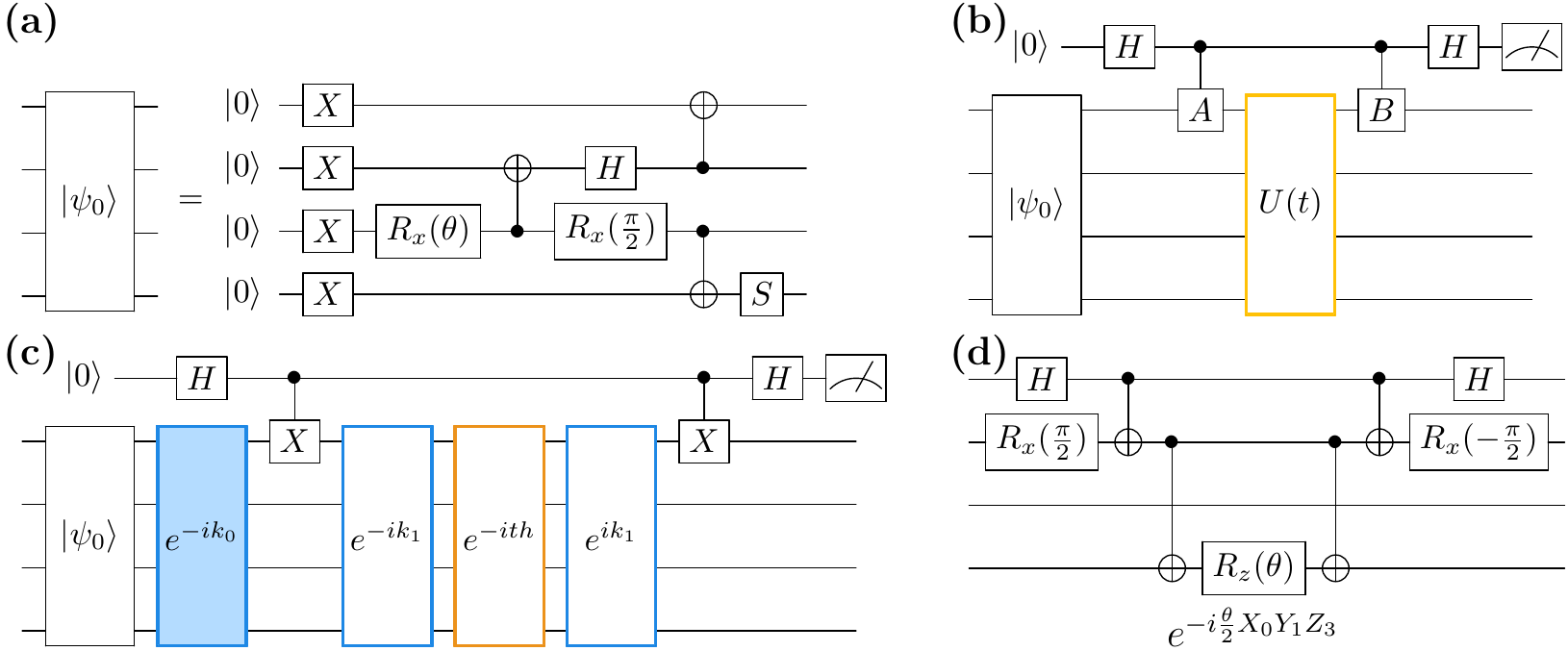}
\caption{(a) Ansatz circuit used to prepare the ground state. (b) General Hadamard interference type circuit used to compute $\Re[\ev{B(t)A}]$. (c) Block decomposed Green's function circuit used in the final computation. The property $[k_0, X_0] = 0$ allows for commuting the $k_0$-block through the CNOT (i.e., controlled-$X$) gate so it need only be applied once. (d) A general circuit showing the implementation of the exponential of a Pauli string, served as a template to decompose the blocks (such as $e^{-ik_1}$ and $e^{-ith}$) in panel (c).}
  \label{fig:Circuit Elements}
\end{figure*}

\section{Hardware implementation} \label{sec:hardwareImpl}
The general circuit used to evaluate Green's function is constructed using three major components: the initial ground state preparation, the time evolution, and the Hadamard-test measurement. Specifically, the system qubits must be first instantiated in the entangled ground state $\ket{\psi_0}$ of the $\hat{H}_\text{AIM}$ with the given $U$ of the Hubbard model and the current $V$ of the DMFT iteration loop. The Green's function expectation is then evaluated using measurements on an ancilla qubit introduced through a Hadamard-test-like interference circuit~\cite{kreula_few-qubit_2016, pedernales2014efficient}. The interference circuit allows for the operator $X_0(t) = U^\dagger(t) X_0 U(t)$ in \cref{eq:reducedGreens} to be implemented using only a single instance of a time evolution operator $U(t)$, which is itself implemented using the sequence of Pauli-exponential gates determined using a set of Cartan decomposition solutions $k_0$, $k_1$, and $h$. The circuit structure remains the same since $\mathfrak{g}(\hat{H}_\text{AIM})$ is independent of specific values of $U \neq 0$ and $V \neq 0$, while the phase parameters of the Pauli-exponential gates are updated once per iteration of the DMFT loop when the Hamiltonian is updated with a new value of $V$.

In \cref{sec:hardwareGateError}, we tabulate the qubit coherence time values (\cref{tab:T1T2time}) and the CNOT gate errors (\cref{tab:CNOTerror}) for {\ibmqManila}, which are extracted from the calibration data recorded at the time of quantum simulation.

\subsection{Circuit components} \label{sec:hardwareCirc}
Our ground state for all values of $V$ and $U$ was initialized using the ansatz circuit in \cref{fig:Circuit Elements}(a), constructed from only a single variational parameter $\theta$ and three nearest-neighbor CNOTs, which correspond to the minimum number of variational parameters and CNOTs need to entangle four qubits and encode the dependence on the ratio $V/U$. The ansatz circuit was initially obtained by manually simplifying the UCCSD circuit in Fig.~7 of Ref.~\cite{McCaskey2019} (but with the two single excitation blocks moved to the right end of the circuit and the two rotation angles set to $\pi/2$); our simplified circuit appears similar to (but still simpler than) the symmetry preserving circuit in Fig.~1 of Ref.~\cite{YeterAydeniz2021}. In \cref{sec:gsCircProof} we prove that the ansatz circuit in \cref{fig:Circuit Elements}(a) prepares the exact ground state of AIM Hamiltonian \cref{eq:AIM2Site}. The value $\theta$ is determined by minimizing the energy through a simulated Variational Quantum Eigensolver.

A generic circuit~\cite{kreula_few-qubit_2016, keen_quantum-classical_2020, pedernales2014efficient} used for evaluating the expectation value $\ev{B(t)A}$, e.g., the Green's function in \cref{eq:reducedGreens}, is shown in \cref{fig:Circuit Elements}(b). The real component of the expectation is determined through a measurement on the ancilla qubit: $\Re[\ev{B(t)A}] = \ev{Z_a} = \text{Pr}(0_a) - \text{Pr}(1_a)$. The corresponding imaginary component, which is not required for our purposes, can be evaluated as $\ev{Y_a}$. 

The Cartan decomposition is computed using the Cartan Quantum Synthesizer Python package~\cite{steckmann_cartan_2021}. For a given solution $k = \sum_j \kappa_j k_j$ to the Cartan decomposition and the corresponding element $h = \sum_j \eta_j h_j$, the time evolution operator is implemented using a sequence of single Pauli string exponential of the form in \cref{fig:Circuit Elements}(d): for example, the factors in $e^{-it\sum_j \eta_j h_j} = \prod_j e^{-it\eta_j h_j}$. The decomposition $k = k_0 + k_1$ (\cref{fig:AIM Decomposition})  factorizes $U(t)$ in \cref{fig:Circuit Elements}(b) into the circuit in \cref{fig:Circuit Elements}(c), which follows from commuting $e^{-ik_0}$ through $X_0$ as follows.
\begin{widetext}
\begin{align}
    \ev{U^\dagger(t) X_0 U(t) X_0}
 &= \ev**{\qty(e^{ik_0}e^{ik_1}e^{ ith}e^{-ik_1}e^{-ik_0}) X_0  
          \qty(e^{ik_0}e^{ik_1}e^{-ith}e^{-ik_1}e^{-ik_0}) X_0}
         {\psi_0}  \notag \\ 
 &= \ev**{\qty(e^{ik_1}e^{ ith}e^{-ik_1}) X_0
          \qty(e^{ik_1}e^{-ith}e^{-ik_1}) X_0}
         {e^{-ik_0} \psi_0} 
\end{align} 
\end{widetext}
Instead of the initial state $\ket{\psi_0}$, we prepare $e^{-ik_0}\ket{\psi_0}$ and time evolve using $e^{ik_1} e^{-ith} e^{-ik_1}$. A combination of additional manual and algorithmic transpiling through Qiskit~\cite{Qiskit} reduces the full cost of the final circuit to $77$ nearest-neighbor CNOTs.

\begin{figure*}[t!]
    \centering
    \includegraphics[trim={0.25in 0 0 0},clip,width=\linewidth]{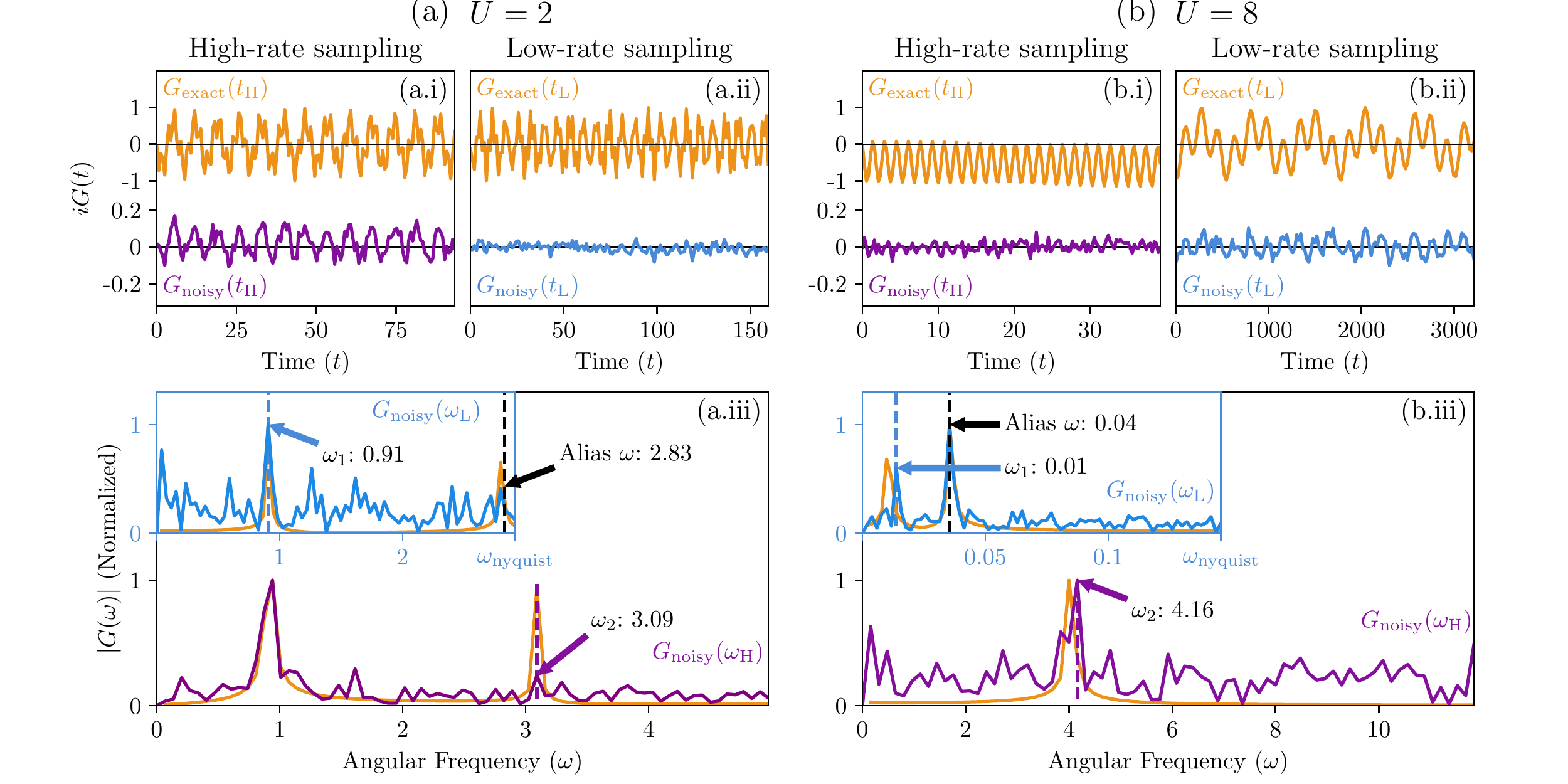}
    \caption{Green's function sampled on the quantum computer {\ibmqManila} at self-consistency. Initial conditions: (a) $U = 2$ and $V_{\text{initial}} = 0.964$ and (b) $U=8$ and $V_{\text{initial}} = 0.119$.  (i/ii) The normalized Green's function with a phase correction (top, shifted vertically) and the actual, noisy results (bottom) with high $(t_H)$ and low $(t_L)$ sampling rates to evaluate the high frequency signal $\omega_2$ and and the low frequency signal $\omega_1$, respectively. (iii) The discrete Fourier transform showing the ideal frequencies (solid, orange) and the evaluated peaks (dashed) for both frequencies. Spurious peaks at $\omega=0$ have been removed. (a) Returns a value of $V_{\textrm{new}} = 0.944$ and (b) returns a value of $V_{\textrm{new}} = 0.116$, both within the tolerance of 0.02.}
    \label{fig:G(t) results}
\end{figure*}

\subsection{Error mitigation} \label{sec:errorMit}
Beyond the noise reductions through careful compilation of the circuit, we implement three methods in an effort to mitigate errors during the \emph{runtime}. First, randomized Cartan solutions are employed in compiling the circuit to mitigate coherent noise, especially the noise due to over rotation of the entangling gates~\cite{zhang_hidden_2021}. The distinct Cartan solutions of $k$ vectors that minimize $f(k)$ are obtained by using different initial conditions in the minimization. We observe that averaging the Green's function measurements from multiple circuits compiled using different $k$ solutions indeed reduces the error in the evaluation. In this work, we use two different solutions to the Cartan decomposition. 

The second method used to reduce error is measurement error mitigation, which serves as an initial step in correcting noise in the experiment results. We process the quantum measurements through the native measurement error mitigation procedure in Qiskit~\cite{Qiskit-Textbook, Qiskit}.

The third method follows from post-selection of the bit strings from measurements on the fermion system qubits, which we find is the most effective among the three error mitigation methods used. The Hadamard test used to measure the Green's function does not require a measurement of these qubits, instead assuming that the system qubits are traced out of the final circuit when measuring $\ev{Z_a}$ on the ancilla qubit. The expectation $\ev{Z_a}$ is unaffected if the partial trace operation is replaced by simultaneous measurements on the system qubits in the computational ($Z$) basis. Since the final state of system qubits is a superposition of $U(t)\ket{\psi_{0}}$ and $X_0U(t) X_0\ket{\psi_0}$ (for $Z_a = \pm 1$, the system state is $[U(t) \pm X_0U(t)X_0] \ket{\psi_0}$), both of which consist of bit strings with the same fixed particle number and total spin $S_z$ as the original state $\ket{\psi_0}$. The initial ground state is known to have two fermions with a total spin $S_z=0$, meaning one particle in each spin sector: for spin $\ua$ ($\da$), bit string $\ket{q_0 q_1}_{\ua} (\ket{q_2 q_3}_{\da}) = \ket{10}$ or $\ket{01}$, so for the evaluation of Green's function by the expectation $\ev{Z_a}$, we only include the shots when the measured bit strings of system qubits satisfy these constraints. This post-selection procedure corresponds to checking for an odd number of bit-flip errors in the fermion system qubits which we expect to affect the final ancilla measurement. On the quantum hardware used in this work, {\ibmqManila}, approximately 65\% of the circuit evaluations are discarded due to this correction, which is applied after all other error mitigation techniques. 

\begin{figure*}[t!]
  \centering
  \includegraphics[trim={0.3in 0 0 2in},clip,width=\linewidth]{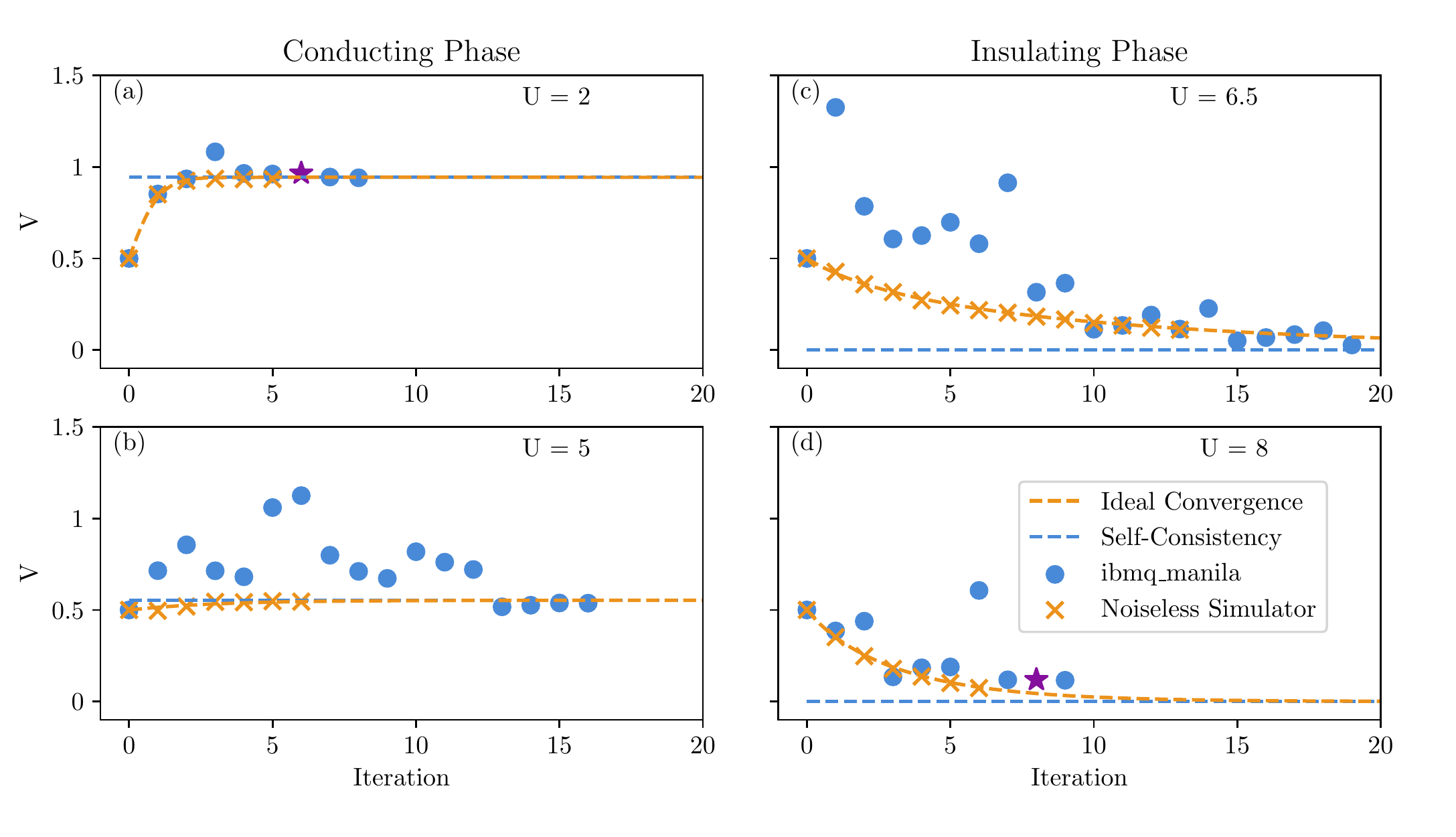}
\caption{DMFT iteration step convergence behavior above [(a) and (b)] and below [(c) and (d)] the critical $U_c = 6$. Despite hardware noise induced error in updating $V$, all converged values are within a stringent small error tolerance we choose, with the except of $U = 6.5$ for which we terminate the iteration after no peaks near $\omega = 0$ were located. The starred points for $U=2$ and $U=8$ correspond to the Green's function evaluations plotted in \cref{fig:G(t) results} (a) and (b), respectively. Computing self-consistency near $U_c = 6$ is cost prohibitive due to critical slowing down in convergence, and therefore the results near $U_c$ are omitted from the phase diagram.}
  \label{fig:Convergence}
\end{figure*}

\subsection{Frequency extraction and self-consistency in DMFT loop}%
\label{sec:freqExtra}
Computing Green's function with adequately converged DMFT loops requires minimizing errors while evaluating a series of discrete time points sufficient to determine both the low frequency signal $\omega_1$ and the high frequency signal $\omega_2$. Generally, these criteria are contradictory to a consideration of minimizing the runtime for the evaluation, since increasing the number of shots, the number of randomized Cartan solutions, and the discrete time points evaluated corresponds to increasing accuracy in convergence but significantly increased runtime. One example of such challenge is as follows.
Above the critical $U_c = 6$, at self-consistency the converged frequencies are $\omega_1 = 0$ with $\alpha_1 = 0$ and $\omega_2 = U/2$ with $\alpha_2 = 1/2$. Consequently, for any discrete time step size $\Delta t_\text{H}$ with a Nyquist frequency $\pi / \Delta t_\text{H}$ above the high frequency $\omega_2 = U/2$, sampling to a sufficiently long time $t_\text{L} \gg t_\text{H}$ to distinguish the low frequency signal $\omega_1$ is prohibitively expensive. For example, finding $\omega_1 = 0.01 \pm 0.005$ with $U = 8$ requires over 5000 evaluations using a sampling rate equal to twice $\omega_2$. Instead, we sample Green's function at two different rates to evaluate $\omega_2$ first and then $\omega_1$. Due to frequency aliasing, the order of the sampling is important. Choosing a low sampling rate to accurately evaluate the low frequency $\omega_1$ with sufficiently long time simulation may result in sampling below the Nyquist rate of $\omega_2$, the high frequency signal. For a given sampling rate $\omega_\text{s}$, the alias frequency $\omega_\text{a}$ within the Nyquist frequency $\omega_\text{s}/2$ can be calculated from the true signal frequency $\omega$ using the following simple formula~\cite{cimbala_2012}:
\begin{align}
    \omega_\text{a} 
  = \abs{\omega - \omega_\text{s} \times
          \nint \qty(\frac{\omega}{\omega_\text{s}})},
  \label{eq:alias}
\end{align}
where $\nint(x) \equiv \ceil{\floor{2x}/2}$ is the (round-half-up) nearest integer to $x$. Thus, we evaluate $\omega_2$ first so the high frequency aliased signal appearing in the low frequency $\omega_1$ sampling regime can be discarded (it nevertheless can be used to check the value of $\omega_2$).

\begin{figure*}[t]
  \centering
  \includegraphics[width=\linewidth]{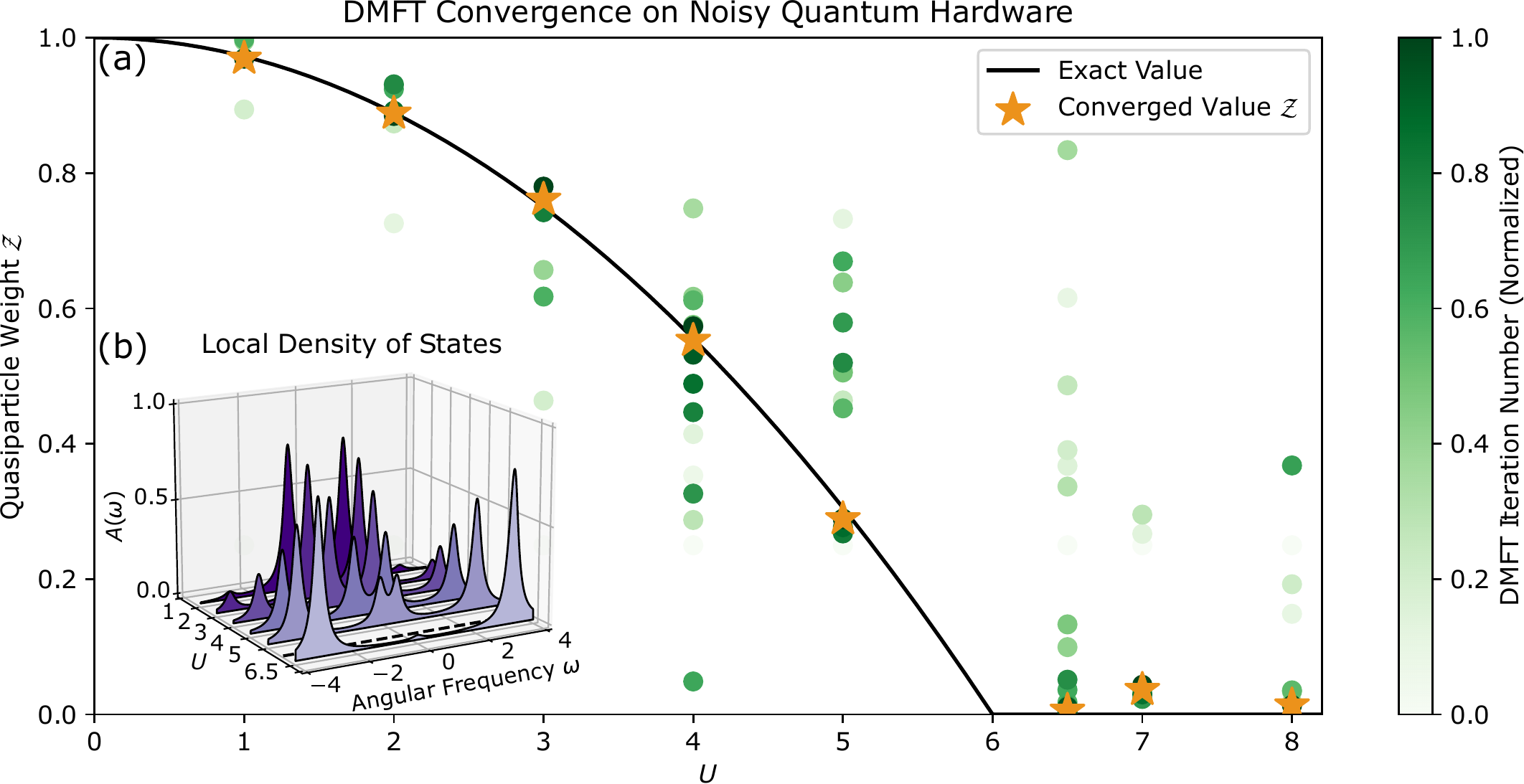}
\caption{(a) Half-filling Hubbard model phase diagram from DMFT computed on a quantum computer. The solid black line corresponds to the exact $\mathcal{Z}$ value of self-consistency given by \cref{eq:selfconsistency}, and the green dots correspond to the computed $\mathcal{Z}$ values in the final few steps of the DMFT loop (the darker the green, the closer it is to the last iteration step). We have omitted from the plot the iteration steps with $\mathcal{Z} > 1$. No data is presented for $5<U<6.5$ because the convergence around the critical $U_c = 6$ slows down significantly resulting cost-prohibitive DMFT loop to reach the convergence. (b) Local density of states $A(\omega)$ above and below $U_c$, where $U_c$ is marked by the dashed line. For visibility, a peak-broadening factor of $\eta = 0.2$ has been included.}
  \label{fig:DMFT_Phase}
\end{figure*}

\Cref{fig:G(t) results} shows the evaluation of $iG(t)$ with ideal simulator in orange (i/ii upper) and the hardware results from high frequency sampling for $\omega_2$ in purple (i, lower) and $\omega_1$ in blue (ii, lower). In each case, $t_\text{H}$ and $t_\text{L}$ are sets of 150 values for time and are chosen with sampling rates between three and ten times greater the than the frequency of $\omega_2$ and $\omega_1$ determined in the previous iteration of the DMFT loop (in both cases above the corresponding Nyquist rate).

To prevent erroneous updates of the loop when an incorrect peak is found due to noise, only the frequency region around an expected peak is searched, as determined by a height criteria based on the most prominent isolated peaks. This process is elaborated on in \cref{sec:peaks}. In the case a condition fails, the particular high frequency or low frequency calculation is rerun until the condition passes before $\mathcal{Z}$ is computed. The loop is iterated until the difference between two sequential results of $V$ is within a tolerance, in our case chosen to be $|\Delta V| \leq 0.02$. The exception is the convergence for $U = 6.5$, in which a prominent peak for $\omega_1$ was not found after 3 attempts and the loop was terminated.

Last, the removal of the dependence on amplitude in computing $\mathcal{Z}$ as in \cref{eq:ZderivativeOmegaOnly} is essential to the success of the DMFT calculation on noisy hardware. The amplitudes of the observed Green's function on hardware are 5 to 15 times lower than in the ideal case, and in general relative signal amplitude $\alpha_1 / \alpha_2$ is not reliably preserved by observed reduced amplitudes.

\section{Results}
Despite significant noise of actual quantum hardware, the quasiparticle and Hubbard band frequencies are preserved in the final discrete Fourier transform of the measured Green's function, allowing for reasonable updates to the DMFT loop, as shown in \cref{fig:Convergence}. The ideal convergence (orange dashed curve in \cref{fig:Convergence}) obtained using the analytical form of the impurity Green's function~\cite{lange_renormalized_1998}, which is interpolated to serve as a guideline, is compared to the convergence of our algorithm executed using the noiseless simulator (orange crosses) and the quantum hardware (blue dots). In the insulating phase [panels (c) and (d)], on the noiseless simulator, our algorithm fails to converge to \emph{exact zero} for $\omega_1$ due to the difficulty to identity the corresponding peak with vanishing amplitude $\alpha_1$. On the other hand, despite significant deviations from the ideal convergence behavior, the convergence on the quantum hardware still trends toward the final self-consistency (blue dashed horizontal line) within a small error tolerance. In addition, in the case of $U = 6.5$, these fluctuating deviations appear to increase the rate of convergence, but generally we have noticed that the deviations prevent ideal, smooth convergence in spite of the significant filtering and error mitigation.
 
\Cref{fig:DMFT_Phase} shows the phase diagram of the quasiparticle weight $\mathcal{Z}_\text{noisy}$ (orange stars) produced on quantum hardware, plotted against the exact solutions for $\mathcal{Z}_\text{exact}$ (black curve)~\cite{potthoff_two-site_2001}, where
\begin{align} \label{eq:selfconsistency}
  \mathcal{Z}_\text{exact} &=
  \begin{cases}
    1 - (U/6)^2, & 0 \leq U < U_c = 6, \\
    0,           & U \geq 6 .
  \end{cases}
\end{align}
The green dots in \cref{fig:DMFT_Phase}(a) are intermediate results obtained in each iteration and the color gradient shows the convergence toward the final value $V_\text{noisy}$, which is taken to be the average of the final two steps in the iteration loop. The inset~\cref{fig:DMFT_Phase}(b) shows the self-consistent local density of states above and below $U_c$.
For $U > U_c$, converged $\alpha_1 = 0$ (and $\mathcal{Z} = 0$) at the self-consistency, requiring a very good signal-to-noise ratio in the results to determine convergence. Therefore, $\mathcal{Z}$ in \cref{fig:DMFT_Phase} for $U > U_c$ saturates at the signal-to-noise floor, which is above zero but small enough to be distinguished from the values of the conducting phase, allowing us to mark the phase transition. In this regime, the fast-forwarding enabled by Cartan decomposition is essential to appropriately study the dynamics over very long times. Results for $U$ very close to $U_c$ are omitted, as critical slowing down confounds the convergence within a reasonable number of iteration steps~\cite{Joo_Oudovenko_2001, Kotliar_Lange_Rozenberg_2000}. 
   
\section{Error Analysis and Future Directions} \label{sec:future_directions}
DMFT remains an impurity-based technique of great interest due to its computational capacity and broad applicability. Despite the promise of asymptotic scaling for Hamiltonian simulation problems in the FTQC regime even the simple case of simulating the two-site DMFT problem has remained intractable on accessible NISQ hardware. This diametrically opposes the platonic ideal of FTQC. In reality, both algorithmic errors and physical errors, from imperfect gates and environmental interactions, arise and must be accounted for. Physical errors severely constrain the long-time simulations, and thus the lowest frequency responses. 

We analyze error scalings in order to understand the competition between physical and algorithmic errors in fast-forwarded quantum computations vs (second-order) Trotter factorizations. We consider the model of $U=2$ near convergence, which has a simulation threshold of $t_{\text{target}} = 8$ required to simulate one full period of $\omega_1$, as informed by~\cref{fig:G(t) results}(a.iii). We then construct a coarse and conservative error model. As the leading cause of physical errors, we assume a CNOT gate fidelity of $\mathcal{F}_\text{CNOT} = 1 - \epsilon_\text{CNOT} = 0.9921$ (as reported by the vendor at the time of data collection) and that all other gates are perfect. Since our proposed algorithm uses 77 CNOTS in this instance for a putative \emph{target} physical fidelity of $\mathcal{F}_\text{runtime} = \mathcal{F}_\text{CNOT}^{77} = 0.543$, although from \cref{fig:G(t) results}(a.iii) one can see that in practice, from the depolarization of the obtained signal, our experimental fidelity $\mathcal{F}_\text{expt} \leq 0.2$. This error model does not capture qualitative physical error details but rather captures the salient qualitative scaling trend. 

\begin{figure}[t]
  \centering
  \includegraphics[trim = 0 0 27 70, clip, width=\columnwidth]{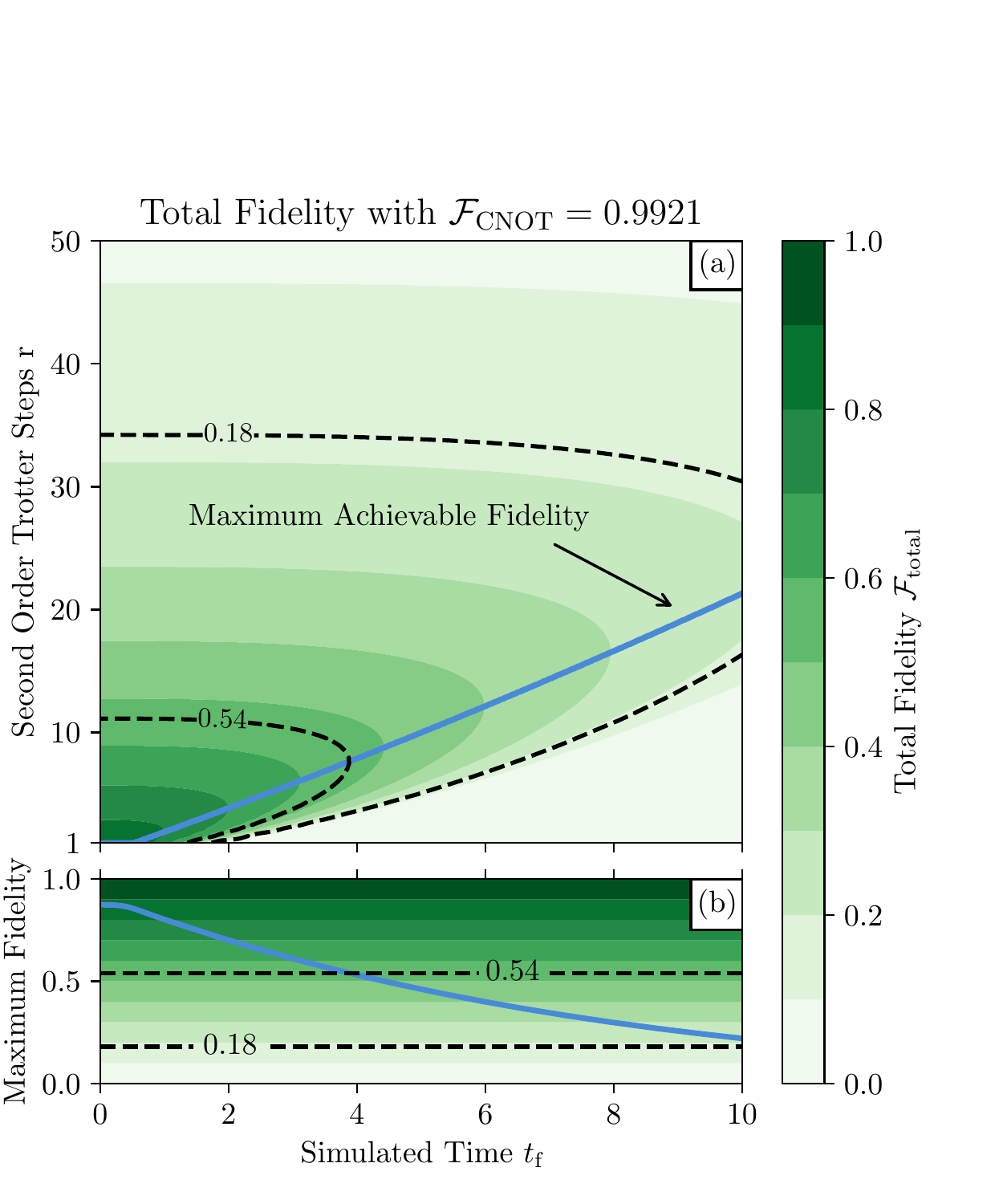}
\caption{(a) Contour plot showing the estimated total fidelity of a second-order Trotter based algorithm for a given simulation time and number of Trotter steps. The dashed lines correspond to the range of estimated total fidelity (0.18--0.54) from our algorithm, while the bold line corresponds to the estimated maximum fidelity that can be achieved for a given target simulation time using the second-order Trotter formula. (b) The line cut corresponding to the bold maximum fidelity line showing the decay of the maximum fidelity as the target time increases. Represents the feasibility of simulating with $U = 2$, $V = 0.94$ near convergence.}
  \label{fig:errors}
\end{figure}

\Cref{fig:errors} illustrates the gate requirements for Trotter-based simulations in terms of a total simulated time ($x$-axis) and the number $r$ of Trotter steps ($y$-axis). The Trotter fidelity $\mathcal{F}_\text{Trotter} = 1 - 0.152t^3/r^2$ is estimated via exact diagonalization with the Frobenius norm in $\|U(t) - V_{\text{Trotter}}(t)\|$ (\cref{sec:error}). The total fidelity is $\mathcal{F}_\text{total} = \mathcal{F}_\text{Trotter} \mathcal{F}_\text{runtime}$.

Building upon the comparison of our algorithm and Trotterization, it is important to note that while asymptotic algorithms such as Trotterization are useful for large-scale system in the long term (fault-tolerance), they are not suitable in the near term due to the accumulation of runtime noise present in current quantum computers. Our algorithm addresses this limitation for the task of solving the two-site DMFT self-consistent impurity Green's function. This work highlights the importance of considering error types and error mitigation on NISQ system. Despite higher initial resource costs (CNOT counts) compared to existing experimental and theoretical works~\cite{keen_quantum-classical_2020, Backes2023, Rungger2019, Jaderberg2020}, our algorithm preserves the frequency signal despite significant depolarizing noise: shifting the error into a single component of the analysis that can be effectively mitigated, which is in contrast to methods such as Trotterization that introduce errors in both the frequency domain (chirping) and amplitude (depolarizing noise). Our error mitigation methods such as Cartan randomized compiling, post-selection of measurement results, and Fourier filtering mitigate the depolarizing noise while increasing the fidelity of the frequency signals.  

Although generally accepted that error mitigation cannot effectively mitigate noise with polynomial overhead~\cite{Quek2022, Chen_Cotler_Huang_Li_2022}, the existence of a quantum advantage from NISQ computers remains open and error mitigation techniques and algorithm design allow will allow increased information processing power from an otherwise noise limited quantum computer. The extent to which error mitigation and error-aware circuits can approach quantum advantage is of significant interest in demonstrating useful, if limited, NISQ quantum applications. Additionally, mitigating and understanding sources and types of runtime or algorithmic errors may allow for reducing the threshold for which fault-tolerant quantum algorithms can be applied~\cite{Suzuki_Endo_Fujii_Tokunaga_2022, Piveteau_Sutter_Bravyi_Gambetta_Temme_2021} by lowering the target total fidelity at the cost of additional sampling overhead or runtime overhead.
\section{Conclusion}
We have demonstrated a two-site DMFT calculation on current generation superconducting quantum hardware with linear CNOT connectivity. Compared to previous methods using the Trotter product formula~\cite{keen_quantum-classical_2020} and the variational method~\cite{Rungger2019} that fail to converge in either the conducting (small $U$) or the insulating (large $U$) phase, our work is the first general implementation to obtain converged physical observables across the full DMFT phase diagram over a wide range of $U$. We find that the bottlenecks in the calculation are the noise in the quantum computer and slow convergence near the transition point. To circumvent these issues we introduced a variety of optimization and error mitigation methods including randomized Cartan solutions in the time evolution, measurement error mitigation, analysis of alias signals in the DFT, and post selection of data. The post selection of data includes enforcement of particle number and total spin conservation since the Hamiltonian under consideration cannot create/destroy particles or flip the spin of particles. 

As demonstrated in this work, Cartan fast-forwarding serves to encode and exploit frequency information despite significant noisy operations on the quantum computer. Although the algorithm used scales poorly with the number of lattice sites in models of \emph{interacting} fermions---here for the four spin-orbital simulation the fixed depth of the Cartan algorithm is significantly longer than a single Trotter step---it provides access to simulations over much longer time scales when the full Trotter circuit depth eventually overtakes the fixed depth of the Cartan algorithm. Thus, for calculations which depend on oscillation frequencies, such as the DMFT and other embedding problems, this and other fast-forwarding algorithms may prove valuable in in the near term, especially when tailored for hardware connectivity.

\begin{acknowledgments}
T.S. was supported in part by the U.S. Department of Energy, Office of Science, Office of Workforce Development for Teachers and Scientists (WDTS) under the Science Undergraduate Laboratory Internship program. T.K. is supported by the U.S.~Department of Energy, Office of Science, Office of Workforce Development for Teachers and Scientists, Office of Science Graduate Student Research (SCGSR) program. The SCGSR program is administered by the Oak Ridge Institute for Science and Education (ORISE) for the DOE. ORISE is managed by ORAU under contract number DE-SC0014664. A.F.K. was supported by the Department of Energy, Office of Basic Energy Sciences, Division of Materials Sciences and Engineering under Grant No. DE-SC0019469. E.F.D. acknowledges DOE ASCR funding under the Quantum Computing Application Teams program, FWP number ERKJ347. Y.W. acknowledges DOE ASCR funding under the Quantum Application Teams program, FWP number ERKJ335. This research used resources of the Oak Ridge Leadership Computing Facility. Access to the IBM Q Network was obtained through the IBM Q Hub at NC State.  
\end{acknowledgments}

\section*{Data availability}
The data and code that support the findings in this study are available at \url{https://doi.org/10.5281/zenodo.5914669}

\onecolumngrid
\appendix

\section{Jordan-Wigner Transform} \label{sec:JW}
For digital quantum simulation of fermionic systems, generally we algebraically map the fermion creation and annihilation operators for $n$ fermion modes to the $n$-qubit Pauli string operators $\bigotimes\limits_{l=0}^{n-1} \hat{\sigma}^{a_l}$ ($a_l \in \{x, y, z, 0\}$), which are Kronecker tensor products of the Pauli matrices $\hat{\sigma}^x = X = \smqty(\pmat{1})$, $\hat{\sigma}^y = Y = \smqty(\pmat{2})$, and $\hat{\sigma}^z = Z = \smqty(\pmat{3})$ and the identity matrix $\hat{\sigma}^0 = I = \smqty(\pmat{0})$. We introduce the notation $\hat{\sigma}_l^{a} = I^{\otimes (l-1)} \otimes \hat{\sigma}^{a} \otimes I^{\otimes (n - l)}$ for the type of Pauli strings acting on only a single qubit. For example, $Z_l = \hat{\sigma}_l^{z} = I^{\otimes l} \otimes Z \otimes I^{\otimes (n - 1 - l)}$. The Jordan-Wigner transform~\cite{nielsen_fermionic_nodate, Jordan_Wigner_1928, Lieb_Schultz_Mattis_1961} used in this work provides one such mapping between the fermionic and qubit algebra. The details of the Jordan-Wigner transform are as follows.

First, map the indexed fermionic states in the Fock basis to indexed qubit states in the computational basis. We employ a mapping where the occupancy numbers of spin-$\ua$ modes are enumerated first followed by that of spin-$\da$ modes so that the Fock basis and computational basis states are both given by the same ``bit string'' $\ket{n_{0\ua}, n_{1\ua}, \cdots, n_{0\da}, n_{1\da}, \cdots}$, where $n_{j\sigma} = 1$ ($n_{j\sigma} = 0$) for occupied (unoccupied) modes of spin-$\sigma$ and lattice site $j$.
    
Second, map the indexed fermionic operators (site index $0\leq j\leq N_b$) to the corresponding qubit operators as follows.
\begin{subequations}
  \label{eq:JWT}
\begin{alignat}{2}
  \hat{c}^\pdag_{j\ua} &= \frac{1}{2} Z_0 \cdots Z_{j-1} (X_j + iY_j),\ 
 &\hat{c}^\dagger_{j\ua} &= \frac{1}{2} Z_0 \cdots Z_{j-1} (X_j - iY_j), 
  \label{eq:JWTa}
\displaybreak[1] \\
  \hat{c}^\pdag_{j\da} &= \frac{1}{2} Z_0 \cdots Z_{N_b + j} (X_{N_b + 1 + j} + iY_{N_b + 1 + j}),\ 
 &\hat{c}^\dagger_{j\da} &= \frac{1}{2} Z_0 \cdots Z_{N_b + j} (X_{N_b + 1 + j} - iY_{N_b + 1 + j}),
  \label{eq:JWTb}
\displaybreak[1] \\
  \hat{n}_{0\ua} &= \frac{1}{2}(I_0 - Z_0),\ 
 &\hat{n}_{0\da} &= \frac{1}{2}(I_{N_b + 1} - Z_{N_b + 1}).
  \label{eq:JWTc}
\end{alignat}
\end{subequations}
In the above, $Z_{j}\cdots Z_{j'} = \prod_{k=j}^{j'}Z_k$ for $j'\geq j$ and $Z_{j}\cdots Z_{j'} = 1$ for $j' < j$. This mapping preserves the canonical anticommutation relations between fermion operators, which are $\left\{ \hat{c}_{j\sigma}^\pdag , \hat{c}_{j'\sigma'}^\dagger \right\} = \delta_{jj'} \delta_{\sigma\sigma'}$ and $\left\{ \hat{c}_{j\sigma}^\pdag , \hat{c}_{j'\sigma'}^\pdag \right\} = \left\{ \hat{c}_{j\sigma}^\dagger , \hat{c}_{j'\sigma'}^\dagger \right\} = 0$.

Finally, plugging in \cref{eq:JWTa,eq:JWTb,eq:JWTc} to \cref{eq:AIMham} gives
\begin{align}
\begin{split}
  \hat{H}_\text{AIM} = &
  \sum_{i = 1}^{N_b}\frac{V_i}{2}
    \big(
      X_0Z_1 \cdots Z_{i-1}X_i + 
      Y_0 Z_1 \cdots Z_{i-1}Y_i +{} \\
    & X_{N_b+1} Z_{N_b+2} \cdots Z_{N_b+i} X_{N_b+1+i} + 
      Y_{N_b+1} Z_{N_b+2} \cdots Z_{N_b+i} Y_{N_b+1+i}
     \big) \\ 
     &{}+
     \frac{U}{4}(Z_0Z_{N_b+1} - Z_0 - Z_{N_b+1}) -
       \sum_{i = 0}^{N_b} \frac{\epsilon_i - \mu}{2}  (Z_i + Z_{N_b + 1 + i}).
\end{split}
  \label{eq:AIMJWham}
\end{align}
We have dropped a constant term $\frac{U}{4}I_0 I_{N_b + 1} = \frac{U}{4}$ from the above Hamiltonian since a constant energy shift does not affect the dynamics of a system.

For the two-site case, with the impurity site and only one bath site ($N_b=1$), the Hamiltonian \cref{eq:AIMJWham} simplifies significantly. Further, at the half-filling (total two particles in the two-site case), $\mu = \frac{U}{2}$ $\epsilon_0 = 0$, and $\epsilon_1 = \frac{U}{2}$ ~\cite{potthoff_two-site_2001, kreula_few-qubit_2016}. Therefore, the two-site $\hat{H}_\text{AIM}$ with a half-filling ground state is given by \cref{eq:AIM2Site}.

\section{Green's Function Evaluation} \label{sec:GFeval}
Plugging in \cref{eq:JWTa,eq:JWTb,eq:JWTc} to \cref{eq:Gimpta} (with $t'=0$), we obtain
\begin{align}
     \ev**{
       \hat{c}_{0}^\pdag (t)
       \hat{c}_{0}^\dagger}
     {\psi_0}
  &= \mel**{\psi_0}
           {U^\dagger(t) \frac{1}{2} (X_0 + iY_0) U(t) \frac{1}{2} (X_0 - iY_0)}
           {\psi_0}  
\displaybreak[1] \notag \\
  &= \frac{1}{4} \qty[
     \ev{U^\dagger(t) X_0  U(t) X_0} +
   i \ev{U^\dagger(t) Y_0  U(t) X_0} -
   i \ev{U^\dagger(t) X_0  U(t) Y_0} +
     \ev{U^\dagger(t) Y_0  U(t) Y_0} ],
\displaybreak[1] \notag \\
     \ev**{
       \hat{c}_{0}^\dagger
       \hat{c}_{0}^\pdag (t)}
     {\psi_0}
  &= \mel**{\psi_0}
          {\frac{1}{2} (X_0 - iY_0) U^\dagger(t) \frac{1}{2} (X_0 + iY_0) U(t)}
          {\psi_0} 
\displaybreak[1] \notag \\
  &= \frac{1}{4} \qty[
     \ev{X_0 U^\dagger(t) X_0  U(t)} +
    i\ev{X_0 U^\dagger(t) Y_0  U(t)} -
    i\ev{Y_0 U^\dagger(t) X_0  U(t)} +
     \ev{Y_0 U^\dagger(t) Y_0  U(t)} ],
\displaybreak[1] \notag \\
     4i G^R_\text{imp}(t>0)
  &= 4 \ev**{
          \hat{c}_{0}^\pdag (t)
          \hat{c}_{0}^\dagger +
          \hat{c}_{0}^\dagger
          \hat{c}_{0}^\pdag (t)}
        {\psi_0} 
  \notag \\        
  &= \ev{X_0(t) X_0} + i\ev{Y_0(t) X_0} 
     - i\ev{X_0(t) Y_0} + \ev{Y_0 (t) Y_0} + {}
  \notag \\
  &\mathrel{\phantom{=}}
     \ev{X_0 X_0(t)} + i\ev{X_0 Y_0(t)}
     - i\ev{Y_0 X_0(t)} + \ev{Y_0 Y_0 (t)},
  \label{eq:fullGreens}
\end{align}
where $X_0(t) \equiv U^\dagger(t) X_0  U(t)$, $Y_0(t) \equiv U^\dagger(t) Y_0  U(t)$, $\ev*{\hat{O}} \equiv \ev{\hat{O}}{\psi_0}$. Measuring the 8 terms in function $G^R_\text{imp}(t)$ would require 16 total circuits: two circuits per term for the real and imaginary components, respectively. Using certain symmetries of the impurity Hamiltonian and the ground state we can show that
\begin{subequations}
\begin{alignat}{2}
    \ev{Y_0(t)Y_0} &= \ev{X_0(t)X_0}, \
  & \ev{Y_0Y_0(t)} &= \ev{X_0X_0(t)}.
  \label{eq:Rzsym} \\
    \ev{Y_0(t)X_0} &= \ev{Y_0X_0(t)}, \
  & \ev{X_0(t)Y_0} &= \ev{X_0Y_0(t)}.
  \label{eq:Tsym} 
\end{alignat}
\end{subequations}
Using \cref{eq:Rzsym,eq:Tsym}, we find $4i G^\text{R}_\text{imp}(t>0) = 2[\ev{X_0(t) X_0} + \ev{X_0 X_0(t)}] = 2[\ev{X_0(t) X_0} + \ev{X_0(t) X_0}^*] = 4\Re \ev{X_0(t) X_0}$, which gives \cref{eq:reducedGreens} $i G^\text{R}_\text{imp}(t>0) = \Re \ev{X_0(t) X_0}$. This reduces the Green's function evaluation to a single measurement circuit for $\Re \ev{X_0(t) X_0}$.

Now we prove \cref{eq:Rzsym} first. The impurity Hamiltonian $\hat{H}_\text{AIM}$ given by \cref{eq:AIM2Site} is invariant under rotation $R_{z,01} = e^{-i\frac{\pi}{4}(Z_0+Z_1)} = e^{-i\frac{\pi}{4}Z_0} e^{-i\frac{\pi}{4}Z_1} $, i.e., $R_{z,01} \hat{H}_\text{AIM} R_{z,01}^\dagger = \hat{H}_\text{AIM}$, so $[\hat{H}_\text{AIM}, R_{z,01}] = [\hat{H}_\text{AIM}, R_{z,01}^\dagger] =0$ and $[U(t), R_{z,01}] = [U^\dagger(t), R_{z,01}] = 0$. Since the ground state $\ket{\psi_0}$ of $\hat{H}_\text{AIM}$ is not degenerate and $[\hat{H}_\text{AIM}, R_{z,01}]=0$, $\ket{\psi_0}$ must be the eigenstate of the unitary operator $R_{z,01}$. Therefore, $R_{z,01} \ket{\psi_0} = e^{i\phi} \ket{\psi_0}$, $R_{z,01}^\dagger \ket{\psi_0} = e^{-i\phi}  R_{z,01}^\dagger e^{i\phi} \ket{\psi_0} = e^{-i\phi} R_{z,01}^\dagger R_{z,01} \ket{\psi_0} = e^{-i\phi} \ket{\psi_0}$, and $\bra{\psi_0} R_{z,01} = \bra{\psi_0} e^{i\phi}$. Now we can prove, for example, $\ev{Y_0(t)Y_0} = \ev{X_0(t)X_0}$.
\begin{align*}
     \ev{Y_0(t)Y_0}
  &= \ev{U^\dagger(t) Y_0 U(t) Y_0}{\psi_0}
\displaybreak[1] \\  
  &= \ev{U^\dagger(t)
         R_{z,01} X_0 R_{z,01}^\dagger
         U(t)
         R_{z,01} X_0 R_{z,01}^\dagger}
     {\psi_0} 
\displaybreak[1] \\
  &= \ev{R_{z,01} U^\dagger(t)
         X_0 R_{z,01}^\dagger R_{z,01}
         U(t)
         X_0 R_{z,01}^\dagger}
     {\psi_0}
\displaybreak[1] \\     
  &= \ev{U^\dagger(t) X_0 U(t) X_0}{\psi_0}
   = \ev{X_0(t)X_0}.
\end{align*}
Similarly, we can prove $\ev{Y_0Y_0(t)} = \ev{X_0X_0(t)}$. 

To prove \cref{eq:Tsym}, we use the time-reversal symmetry of the Hamiltonian $\mathcal{T} \hat{H}_\text{AIM} \mathcal{T}^{-1} = \hat{H}_\text{AIM}$. The time-reversal symmetry operator $\mathcal{T} = e^{-i\frac{\pi}{2} (Y_0 + Y_1 + Y_2 + Y_3)} \mathcal{K} = Y_{0}Y_{1}Y_{2}Y_{3} \mathcal{K}$, where the operator $\mathcal{K}$ takes the complex conjugation. $\mathcal{T} X_0 \mathcal{T}^{-1} = -X_0$ due to $\mathcal{K} X_0 = X_0$ and $Y_0 X_0 = -X_0 Y_0$. $\mathcal{T} Y_0 \mathcal{T}^{-1} = -Y_0$ due to $\mathcal{K} Y_0 = -Y_0$. Similar to $R_{z,01}$ symmetry operator, we have $\mathcal{T} \ket{\psi_0} = \ket{\psi_0}$ and $\mathcal{T} U(t) \mathcal{T}^{-1} = U(-t)$. Beginning with the time-translation invariance result $\ev{Y_0(t)X_0} = \ev{Y_0X_0(-t)}$, we prove $\ev{Y_0(t)X_0} = \ev{Y_0X_0(t)}$ as follows.
\begin{align*}
     \ev{Y_0(t)X_0}
  &= \ev{Y_0X_0(-t)}
   = \ev{Y_0 U(t) X_0 U(-t)}{\psi_0}
\displaybreak[1] \\  
  &= \ev{\mathcal{T}^{-1}
         (\mathcal{T} Y_0 \mathcal{T}^{-1})
         (\mathcal{T} U(t) \mathcal{T}^{-1})
         (\mathcal{T} X_0 \mathcal{T}^{-1})
         (\mathcal{T} U(-t) \mathcal{T}^{-1})
         \mathcal{T}}
     {\psi_0} 
\displaybreak[1] \\
  &= \ev{Y_0 U(-t) X_0 U(t)}
     {\psi_0}
\displaybreak[1] \\     
  &= \ev{Y_0 U^\dagger(t) X_0 U(t)}
     {\psi_0}
   = \ev{Y_0X_0(t)}.
\end{align*}
Similarly, we can prove $\ev{X_0(t)Y_0} = \ev{X_0Y_0(t)}$.

\section{Self-Energy and Its Derivative at Zero Frequency}
\label{sec:Self-energy-derivative}
We drop the impurity ``imp'' subscript below for simplicity. The retarded interacting impurity Green's function of the two-site AIM in the time domain has the form
\begin{align}
     iG(t>0) 
  &= \alpha_1 e^{ i\omega_1 t} +
     \alpha_1 e^{-i\omega_1 t} +
     \alpha_2 e^{ i\omega_2 t} +
     \alpha_2 e^{-i\omega_2 t}
   = 2 \alpha_1 \cos(\omega_1 t) +
     2 \alpha_2 \cos(\omega_2 t), \label{eq:Gt}
\end{align}
where $2\alpha_1 + 2\alpha_2 = 1$ due to the spectral function sum rule. To extract the poles $\omega_{1,2}$ and amplitudes $\alpha_{1,2}$, instead of fitting the sampled time series with the above function form~\cite{kreula_few-qubit_2016, keen_quantum-classical_2020, Jaderberg2020}, we obtain $\omega_{1,2}$ directly from the discrete Fourier transform of the data, according to the analytic Fourier transform of \cref{eq:Gt} given by
\begin{align}
    G(\omega) 
  &= \frac{\alpha_1}{\omega - \omega_1} +
     \frac{\alpha_1}{\omega + \omega_1}  +
     \frac{\alpha_2}{\omega - \omega_2}  +
     \frac{\alpha_2}{\omega + \omega_2}. \label{eq:Gw}
\end{align}


The noninteracting Green's function is given by $G_0(\omega) = \frac{1/2}{\omega - V} + \frac{1/2}{\omega + V}$. The self-energy is then given by Dyson's equation as follows. 
\begin{align}
     \Sigma(\omega) 
  &= \frac{1}{G_0(\omega)} - \frac{1}{G(\omega)} \label{eq:SigmaDyson}\\   
  &= \frac{\omega^2 - V^2}{\omega} -
     \frac{(\omega^2 - \omega_1^2)(\omega^2 - \omega_2^2)}
          {2\omega [\omega^2(\alpha_1 + \alpha_2) - (\alpha_1 \omega_2^2 + \alpha_2 \omega_1^2)]}. \label{eq:SigmaFull}
\end{align}
Since $G_0(0) = G(0) = 0$, if the divergences of $G_0^{-1}(\omega)$ and $G^{-1}(\omega)$ do not cancel exactly at $\omega = 0$, $\Sigma(\omega)$ diverges, resulting unphysical pole and contradicting to the expected behavior of self-energy before phase transition happens. So the physically meaningful self-energy must satisfy the following constraint:
\begin{align}
  & \lim_{\omega\to 0} \qty[\omega\Sigma(\omega)] = 0
\displaybreak[1] \notag \\  
  \implies & \lim_{\omega\to 0} \qty[ \omega^2 - V^2 -
     \frac{(\omega^2 - \omega_1^2)(\omega^2 - \omega_2^2)}
          {2[\omega^2(\alpha_1 + \alpha_2) - (\alpha_1 \omega_2^2 + \alpha_2 \omega_1^2)]} ] = 0
\displaybreak[1] \notag \\  
  \implies & \frac{\omega_1^2 \omega_2^2}{2(\alpha_1 \omega_2^2 + \alpha_2 \omega_1^2)} = V^2.
  \label{eq:SigmaConstr}
\end{align}

The constraint \cref{eq:SigmaConstr} must hold \emph{exactly} for the self-energy to avoid any unphysical pole at $\omega = 0$, which is usually not the case with $\omega_{1,2}$ and $\alpha_{1,2}$ extracted from noisy data. To regularize $\Sigma(\omega)$ at $\omega = 0$ we impose \cref{eq:SigmaConstr} directly, plug it in \cref{eq:SigmaFull} for the self-energy, and then compute the derivative of the self-energy at $\omega = 0$ as follows.
\begin{align}
     \Sigma(\omega) 
  &= \omega - \frac{V^2}{\omega} -
     \frac{(\omega^2 - \omega_1^2)(\omega^2 - \omega_2^2)}
          {2\omega [\omega^2(\alpha_1 + \alpha_2) - \omega_1^2 \omega_2^2/(2V^2)]}
\displaybreak[1] \notag \\ 
  &= \omega -      
     \frac{\omega V^2[\omega^2 + 2V^2(\alpha_1 + \alpha_2) - (\omega_1^2 + \omega_2^2)]}
          {2\omega^2 V^2 (\alpha_1 + \alpha_2) - \omega_1^2 \omega_2^2}.
\\ 
     \left.\dv{\Sigma(\omega)}{\omega}\right|_{\omega = 0}
  &= \lim_{\omega\to 0} \frac{\Sigma(\omega)}{\omega}
\displaybreak[1] \notag \\   
  &= 1 - \frac{V^2[(\omega_1^2 + \omega_2^2) - 2V^2(\alpha_1 + \alpha_2)]}{\omega_1^2 \omega_2^2}.
\end{align}

Finally, using the sum rule $2\alpha_1 + 2\alpha_2 = 1$, we obtain the \cref{eq:ZderivativeOmegaOnly} used in the main text to compute the quasiparticle weight.
\begin{align}
    \mathcal{Z} 
 &= \flatfrac{1}{\left[ 1 - 
      \dv{{\Re}\Sigma(\omega)}{\omega} \right]_{\omega=0}}
  = \frac{\omega_1^2 \omega_2^2}{V^2(\omega_1^2 + \omega_2^2 - V^2)}.
\end{align}

We remark here that in addition to plotting $\mathcal{Z}$ in \cref{fig:DMFT_Phase}(a), we have also plotted in the inset~\cref{fig:DMFT_Phase}(b) the local density of states $A(\omega) = -\frac{1}{\pi} \Im G(\omega + i\eta)$, where $G(\omega + i\eta)$ is given by \cref{eq:Gw} with the small imaginary part $\eta = 0.2$ and $\alpha_{1,2}$ computed using the extracted frequencies $\omega_{1,2}$ as follows [obtained from the sum rule and the \cref{eq:SigmaConstr}].
\begin{align}
  \alpha_2 = \frac{(\omega_1/V)^2 - 1}{2[(\omega_1/\omega_2)^2 - 1]}, \quad
  \alpha_1 = \frac{1}{2} - \alpha_2.
\end{align}

  \mycomment{
From the definitions of the self-energy and its derivative, we find:
\begin{align}
     \Sigma(\omega) 
  &= \frac{1}{G^{(0)}(\omega)} - \frac{1}{G(\omega)},
\displaybreak[1] \label{eq:SigmaDyson}\\   
     \dv{\Sigma(\omega)}{\omega}
  &= -\frac{\dv*{G^{(0)}(\omega)}{\omega}}{G_0^2(\omega)}
     +\frac{\dv*{G(\omega)}{\omega}}{G^2(\omega)}
\displaybreak[1] \notag\\
  &= -\frac{1}{\omega^2}\left\{ 
     \frac{\alpha_1(\omega^2 - \omega_2^2)^2\omega_1^2 +
           \alpha_2(\omega^2 - \omega_1^2)^2\omega_2^2}
          {2[\alpha_1(\omega^2 - \omega_2^2) + 
           \alpha_2(\omega^2 - \omega_1^2)]^2}    
     -V^2\right\} + 1 -
     \frac{\alpha_1(\omega^2 - \omega_2^2)^2 +
           \alpha_2(\omega^2 - \omega_1^2)^2}
          {2[\alpha_1(\omega^2 - \omega_2^2) + 
           \alpha_2(\omega^2 - \omega_1^2)]^2} 
\label{eq:dSigma}\\           
  &\equiv -\frac{1}{x}
    \left[\frac{f(x)}{g(x)} - V^2\right] + h(x).
\end{align}
In the last step, we have defined functions $f(x) = \alpha_1(x - x_2)^2 x_1 +\alpha_2(x - x_1)^2 x_2$, $g(x) = 2[\alpha_1(x - x_2) + \alpha_2(x - x_1)]^2$, and $h(x) = 1 - [\alpha_1(x - x_2)^2 +\alpha_2(x - x_1)^2]/g(x)$, where $x = \omega^2$, $x_1 = \omega_1^2$, and $x_2 = \omega_2^2$, for the discussion below. 

Now we derive a stable formula to calculate $\lim\limits_{\omega \to 0} \dv{\Sigma(\omega)}{\omega} \equiv \Sigma'(0)$.
First, $\left. h(x)\right|_{x = \omega^2 = 0}$ is finite because if the denominator of the second term in $h(x)$ is zero at $x = \omega^2 = 0$, this implies that $\alpha_1\omega_2^2 + \alpha_2\omega_1^2 = 0$, which is impossible since $\alpha_1,\ \alpha_2,\ \omega_1,\ \omega_2 \geq 0$, $\alpha_1 + \alpha_2 > 0$, and $\omega_1 + \omega_2 > 0$. Therefore, the first term of $\Sigma'(\omega)$, $-\frac{1}{x}\left[\frac{f(x)}{g(x)} - V^2\right]$, must also be finite as $x=\omega^2 \to 0$. This is only possible if $\left.\frac{f(x)}{g(x)} - V^2\right|_{x=0} = 0$, which gives
\begin{align}
     \left.\frac{f(x)}{g(x)}\right|_{x=0}
  &= \frac{f(0)}{g(0)}
   = \frac{\alpha_1 x_2^2 x_1 + \alpha_2 x_1^2 x_2}
         {2(\alpha_1 x_2 + \alpha_2 x_1)^2}
   = \frac{x_1 x_2 (\alpha_1 x_2 + \alpha_2 x_1)}
         {2(\alpha_1 x_2 + \alpha_2 x_1)^2}
   =\frac{x_1 x_2 }{2(\alpha_1 x_2 + \alpha_2 x_1)} = V^2.
   \label{eq:foverg}
\end{align}

Therefore, we can evaluate the $x\to 0$ limit of the first term $-\frac{1}{x}\left[\frac{f(x)}{g(x)} - V^2\right]$ with L'H\^{o}pital's rule as follows.
\begin{align}
\lim_{x\to 0} -\frac{1}{x}\left[\frac{f(x)}{g(x)} - V^2\right]
  &= \frac{g'(0)V^2 - f'(0)}{g(0)}
\notag \\
  &= \frac{-4V^2(\alpha_1 x_2 + \alpha_2 x_1) (\alpha_1 + \alpha_2) 
           +4V^2(\alpha_1 x_2 + \alpha_2 x_1) (\alpha_1 + \alpha_2)}{g(0)}
   = 0. \label{eq:dSigmaTerm1}
\end{align}

Since the first term of $\Sigma'(\omega)$, \cref{eq:dSigmaTerm1}, is zero, we find
\begin{align}
     &\left.\dv{\Sigma(\omega)}{\omega}\right|_{\omega = 0}
   = h(0)
   = 1 - \frac{\alpha_1 x_2^2 + \alpha_2 x_1 ^2}
              {2(\alpha_1 x_2 + \alpha_2 x_1)^2},
  \label{eq:dSigmaw0a}
\\
     &\mathcal{Z} 
  = \flatfrac{1}{\left[ 1 - 
      \dv{{\Re}\Sigma(\omega)}{\omega} \right]_{\omega=0}}
  = \frac{2(\alpha_1 x_2 + \alpha_2 x_1)^2}
         {\alpha_1 x_2^2 + \alpha_2 x_1 ^2},
  \label{eq:ZdSigmaw0a}
\\
  &x_1 = \omega_1^2,\ x_2 = \omega_2^2. \notag
\end{align}
\Cref{eq:dSigmaw0a,,eq:ZdSigmaw0a} are more stable and robust against numerical or (quantum) simulation error. The stability comes from the removal of the first term in \cref{eq:dSigma} which is mathematically singular if the condition \cref{eq:foverg} is not satisfied exactly.

Last, we use the two conditions (derived from the sum rule and \cref{eq:foverg})
\begin{subequations}
\begin{align}
  \alpha_1 + \alpha_2 &= \frac{1}{2} \\
  x_2 \alpha_1 + x_1 \alpha_2 &= \frac{x_1 x_2}{2V^2}
\end{align}
\end{subequations}
to solve for $\alpha_1$ and $\alpha_2$ in terms of $x_1$, $x_2$, and $V$, and then eliminate them from the \cref{eq:ZdSigmaw0a}. The final result is
\begin{align}
    \mathcal{Z}
 &= \frac{x_1 x_2}{V^2(x_1 + x_2 - V^2)}
  = \frac{\omega_1^2 \omega_2^2}{V^2(\omega_1^2 + \omega_2^2 - V^2)}.
  \label{eq:ZdSigmaw0c}  
\end{align}
  }

\section{Quantum Hardware} \label{sec:hardwareGateError}
The quantum hardware used in this work was {\ibmqManila}, a superconducting device with 5 qubits and a linear qubit topology (qubits are sequentially labeled from one end to the other). It is publicly available through the IBM Quantum Experience. The experiment parameters are designed around the open access quantum job submission limits of 5 sets of 75 circuit evaluations of 8,192 shots each. Thus, we used two distinct solutions to the Cartan decomposition which combine to a total shot count of 16,000 at each of 150 time step evaluations. The first set was reserved for just in time measurement error mitigation circuits, of which there are 32 circuits preparing each of the $2^5$ computational basis states. Assuming a correct evaluation, each DMFT Loop requires approximately 36 minutes to execute on the IBM backend, including the Measurement Error Mitigation circuits. In practice, the update failure condition results in repeated calculations and subsequently increased runtimes for each $V$ update. Tables I and II show the qubit coherence times and the entangling gate properties, respectively. The ancilla qubit was placed at index 0. Calibration data pulled from the Qiskit API~\cite{Qiskit}, and averaged by taking the calibration data at 4 points each day between October 10$^\text{th}$ 2021 and November 10$^\text{th}$ 2021.
\begin{table}[hbt]
  \begin{tabular}{c|c|c} \label{tab:single-qubit-errors}
       Qubit Number & T1 ($\mu s$) & T2 ($\mu s$) \\ \hline
0 (Ancilla) & \ \ 146.18 $ \pm $ \ 29.78 \ \ & \ 94.38 $ \pm $ \ 18.50 \\
1           & \ \ 204.64 $ \pm $ \ 47.57 \ \ & \ 83.02 $ \pm $ \ 15.18 \\
2           & \ \ 148.79 $ \pm $ \ 26.79 \ \ & \ 24.30 $ \pm $ \ 2.55 \\
3           & \ \ 157.19 $ \pm $ \ 36.33 \ \ & \ 63.64 $ \pm $ \ 7.83 \\
4           & \ \ 128.96 $ \pm $ \ 24.11 \ \ & \ 42.78 $ \pm $ \ 2.57
  \end{tabular}
  \caption{\label{tab:T1T2time}Average T1 and T2 coherence times for {\ibmqManila}, averaged over the period of time in which runs were executed.}

  \begin{tabular}{c|c|c} \label{tab:multi-qubit-errors}
      Connection & CNOT Error Rate      & Gate Timing (ns) \\ \hline
      0-1 & \ 0.0070 $ \pm $ \ 0.0012 & \ 295.11 $\pm$ 17.78 \\
      1-2 & \ 0.0099 $ \pm $ \ 0.0017 & \ 487.11 $\pm$ 17.78 \\
      2-3 & \ 0.0071 $ \pm $ \ 0.00080& \ 373.33 $\pm$ 17.78 \\
      3-4 & \ 0.0076 $ \pm $ \ 0.0016 & \ 316.44 $\pm$ 17.78
  \end{tabular}
  \caption{\label{tab:CNOTerror} Average CNOT error rates and gate timings on {\ibmqManila}.}
\end{table}

\section{Verification of the Ground State Preparation Ansatz Circuit}
\label{sec:gsCircProof}
Since the ground state has two fermions with a total spin $S_z = 0$, there must be one particle for each spin sector and the Hilbert space $\mathcal{H}_{(N_{\ua}=1, N_{\da}=1)}$ of this symmetry sector is spanned by basis $\{\ket{q_0 q_1}_\ua \otimes \ket{q_2 q_3}_\da \} = \{\ket{10}, \ket{01}\} \otimes \{\ket{10}, \ket{01}\} = \{ \ket{\phi_1} = \ket{1010}, \ket{\phi_2} = \ket{0101}, \ket{\phi_3} = \ket{1001}, \ket{\phi_4} = \ket{0110}\}$. Since $\hat{H}_\text{AIM} \ket{\phi_i} \in \mathcal{H}_{(1,1)}$, we need only solve the ground state within this matrix block $\mathbf{H} = (\bra{\phi_1}, \bra{\phi_2}, \bra{\phi_3}, \bra{\phi_4})^T \hat{H}_\text{AIM} (\ket{\phi_1}, \ket{\phi_2}, \ket{\phi_3}, \ket{\phi_4})$, where the matrix elements $\mathbf{H}_{i,j} = \mel{\phi_i}{\hat{H}_\text{AIM}}{\phi_j}$ which are easily evaluated using the \cref{eq:AIM2Site} and the following basis transfer matrix:
\begin{subequations}
\begin{align}
     \hat{H}_\text{AIM} \ket{\phi_j} 
  &= V(\ket{\phi_3} + \ket{\phi_4}) + \frac{U}{4} \ket{\phi_j}, \quad (j=1,2), \\
     \hat{H}_\text{AIM} \ket{\phi_j} 
  &= V(\ket{\phi_1} + \ket{\phi_2}) - \frac{U}{4} \ket{\phi_j}, \quad (j=3,4). 
\end{align}
\end{subequations}

Noticing the symmetric form of the above equations, we can obtain a block diagonal matrix $\mathbf{H}' = \spmqty{U/4 & 2V \\ 2V &-U/4} \oplus \spmqty{U/4 & \\ &-U/4}$ using the following new basis: 
$\ket{\phi'_1} = (\ket{\phi_{1}} + \ket{\phi_{2}})/\sqrt{2}$,
$\ket{\phi'_2} = (\ket{\phi_{3}} + \ket{\phi_{4}})/\sqrt{2}$,
$\ket{\phi'_3} = (\ket{\phi_{1}} - \ket{\phi_{2}})/\sqrt{2}$,
$\ket{\phi'_4} = (\ket{\phi_{3}} - \ket{\phi_{4}})/\sqrt{2}$. The second block gives the following (eigenvalue, eigenvector) of $\hat{H}_\text{AIM}$: $(U/4, \ket{\phi'_3})$ and $(-U/4, \ket{\phi'_4})$. The first block $\spmqty{U/4 & 2V \\ 2V &-U/4}$ gives the eigenvalues $\pm \sqrt{4V^2 + (U/4)^2}$ and both eigenvectors must have the same wavefunction form that is a linear combination of two basis vectors $\cos\alpha\ket{\phi'_1} + \sin\alpha\ket{\phi'_2}$ (with different $\alpha$ values for different eigenvalues).

Comparing the eigenvalues, we conclude that the \emph{exact} ground state has the eigenenergy $-\sqrt{4V^2 + (U/4)^2}$ and the wavefunction form $\frac{\cos\alpha}{\sqrt{2}} (\ket{1010} + \ket{0101}) + \frac{\sin\alpha}{\sqrt{2}} (\ket{1001} + \ket{0110})$. Now we verify that the ansatz circuit in \cref{fig:Circuit Elements}(a) prepares the exact ground state wavefunction form by applying the gates sequentially on the initial state $\ket{0000}$. For example, after the first four $X$ gates, $\ket{0000} \to \ket{1111}$; after the next gate $R_x(\theta) = e^{-i\frac{\theta}{2} X_2} = \cos\frac{\theta}{2} - i \sin\frac{\theta}{2}X_2$, $\ket{1111} \to \cos\frac{\theta}{2}\ket{1111} - i \sin\frac{\theta}{2}\ket{1101}$; after the next $CX_{2,1}$, the state becomes $\cos\frac{\theta}{2}\ket{1011} - i \sin\frac{\theta}{2}\ket{1101}$; and so on. The state at the end of the ansatz circuit is $\ket{\psi_0} = \frac{\cos(\theta/2 + \pi/4)}{\sqrt{2}} (\ket{1010} + \ket{0101}) + \frac{\sin(\theta/2 + \pi/4)}{\sqrt{2}} (\ket{1001} + \ket{0110})$, which agrees with the aforementioned exact ground state wavefunction form. 

\section{Detection of Frequency-Domain Green's Function Peaks} \label{sec:peaks}
\paragraph{High frequency peaks.} As mentioned in the main text, we first detect the high frequency ($\omega_2$) peak and then the low frequency ($\omega_1$) peak. For all investigated values of $U$, along the course of the DMFT iterations, the amplitudes ($\alpha_2$) of the high frequency ($\omega_2$) peaks remain significant compared to the level of noise. Thus, we enforce a strict criterion on peak height when detecting the correct frequency-domain Green's function peaks from noisy data. This is done by searching for up to two most prominent peaks with an amplitude greater than two standard deviations above the mean signal strength (essentially corresponding to the background noise). The search is initially only limited in a narrow range around the ``expected'' frequency, i.e., a small window centered around the detected $\omega_2$ of the previous iteration. An ``expected'' frequency region for $\omega_1$ is defined similarly and masked off when searching for $\omega_2$. In the event a peak is not located within the largest search area, the iteration is rerun and Green's function is recomputed. When two prominent peaks are detected, the higher amplitude peak is selected. Peak detection is performed using the \verb|SciPy.signal| library function \verb|find_peaks|. 

\paragraph{Low frequency peaks.} In or near the insulating phase, the amplitudes ($\alpha_1$) of the low frequency ($\omega_1$) peaks vanish near convergence. We therefore reduce the peak height requirement of the peak search as the iterations converge. The criteria are listed as follows, in the order of increasing peak height requirement. The later steps are only used if the previous step returns more than two peaks for the given threshold.
\begin{enumerate}[topsep=0pt, partopsep=0pt, itemsep=0pt, parsep=0pt]
    \item Are there either one or two peaks above the average?
    
    \item If no, are there either one or two peaks one standard deviation above the average?
    
    \item If no, are there either one or two peaks two standard deviations above the average?
    
    \item If no, rerun the entire iteration for both $\omega_1$ and $\omega_2$.
\end{enumerate}
If at any point the answer is yes, take the most prominent of the one or two peaks as $\omega_1$. For high amplitudes, lower thresholds will lead to too many peaks, and increased strictness will eliminate theses extraneous peaks. For low amplitudes, the iteration is more likely to fail or choose an extraneous peak by merit of a much lower signal-to-noise ratio.
For all search regions, any aliased frequencies from $\omega_2$ are eliminated from the search. Near search boundaries, extraneous peaks are often detected and eliminated from the result.

\section{Total Fidelity Estimate of Hamiltonian Simulation Algorithms}
\label{sec:error}
The total fidelity $\mathcal{F}_{\text{tot}}$ of two Hamiltonian-based time evolution algorithms, Trotter and Cartan, is modeled as follows. Assuming independent physical and algorithmic errors, the multiplicativity of the fidelity from orthogonal sources is
\begin{align}
     \mathcal{F}_{\text{tot}} 
  &= \mathcal{F}_{\text{alg}} \mathcal{F}_{\text{runtime}},
\end{align}
where the factor $\mathcal{F}_{\text{alg}}$ takes into account of the algorithmic error and $\mathcal{F}_{\text{runtime}}$ is a function of the counts and fidelity of CNOT gates in the circuits used in the runtime.

In the case of the Cartan decomposition circuit, the algorithmic fidelity is very near unity (to within classical machine errors). We estimate a runtime fidelity $\mathcal{F}_{\text{runtime}} = (\mathcal{F}_{\text{CNOT}})^{77}$ as the circuit execution required 77 CNOTs, which dominate the runtime-errors. Although this presents a very simplified error model, it captures the qualitative scaling of the errors accrued while the running algorithm. 

The second-order Trotter-Suzuki formula is given by 
\begin{align}
     U_2(t_f,r)
  &= \qty[ \qty( e^{-i (t_f/2r) H_0    } \cdots e^{-i (t_f/2r) H_{m-1}} )
           \qty( e^{-i (t_f/2r) H_{m-1}} \cdots e^{-i (t_f/2r) H_0    } ) ]^r
  \equiv \qty[ \tilde{U}_{2}(\tau) ]^r,
\end{align}
where $r$ is the number of Trotter steps, $t_f$ is the final simulation time, $\tau = t_f/r$, and $m$ is the number of non-commuting sets of terms in the Hamiltonian. For a system with $m=2$ such as ours, a single step simplifies to
\begin{align}
     \tilde{U}_{2}(\tau)
  &= e^{-i (\tau/2) H_0} e^{-i \tau H_1}e^{-i (\tau/2) H_0}.
\end{align}
Combining multiple second-order Trotter steps further reduces the decomposition to
\begin{align}
     U_2(t_f,r)
  &= e^{i (\tau/2) H_0} \qty( e^{-i \tau H_0}e^{-i \tau H_1} )^r e^{-i (\tau/2) H_0} \\
  &\equiv e^{i (\tau/2) H_0} U_1(t_f, r) e^{-i (\tau/2) H_0},
\end{align}
where the first-order Trotter formula is defined as $U_1 (t_f, r) = (e^{-i \tau H_0} e^{-i \tau H_1})^r$. The above connection between first-order and second-order Trotter formulas is a result specific for $m=2$ \cite{Layden2022}.

Simulating $r$ steps requires the same resources as would first-order Trotterization, plus a single implementation of the first $H_0$ exponential on the right (the last $H_0$ exponential on the left can be absorbed into the last Trotter step). In this analysis, we choose $H_0 = \frac{1}{4}U(Z_0Z_2)$ and $H_1 = \frac{1}{2}V(X_0X_1 + Y_0Y_1 + X_2X_3 + Y_2Y_3)$. The CNOT costs for each step are 2 CNOTs and 4 CNOTs, respectively. Each of the $X_iX_{i+1} + Y_iY_{i+1}$ ($i=0,2$) exponentials can be implemented using only 2 CNOTs by diagonalizing each into single site Z rotations using the Clifford element $(H_{i}H_{i+1})(S^\dag_{i}S^\dag_{i+1})(C_iX_{i+1})(H_i\otimes I_{i+1})$ to diagonalize the terms. These simplifications give a total CNOT count of $6r + 2$. However, there is additional overhead due to two SWAP gates (each using 3 CNOTs) needed to perform the $Z_0Z_2$ exponential on linear connected hardware and the 3 CNOTs included in the VQE ground state preparation. The runtime fidelity is given by $\mathcal{F}_{\text{runtime}} = (\mathcal{F}_{\text{CNOT}})^{(6r + 11)}$ for total $6r + 11$ CNOT gates.

The Trotter error is computed as a numerical fit for the coefficient of the asymptotic leading order error term $\mathcal{O}(t^3/r^2)$ using exact diagonalizaton of the actual unitary evolution operator $U(t)$: $\| U(t) - V(t)\| \approx 0.152 t^3/r^2$, where $\|\cdot\|$ is the Frobenius norm.

The ``Maximum Achievable Fidelity'' curve (solid thick blue line in \cref{fig:errors}) is then given by optimizing $\mathcal{F}_{\text{tot}}$ over $r \in \mathbb{R}$ for a fixed $t_f$.
%

Our choice of the $t_{\text{target}} \approx 8$ threshold is given by the period $T_1$ of the low frequency component $\omega_1 = 0.884$ or $T_1 = 2\pi / \omega_1 = 7.11$ for our choice of parameters.

\twocolumngrid

%


\end{document}